\documentclass[12pt,a4paper,notitlepage]{report}
\usepackage{amsmath}
\usepackage{latexsym}
\usepackage{amsthm}
\usepackage{braket}
\usepackage{amssymb}
\usepackage{graphicx}
\usepackage{subcaption}
\usepackage[pdfstartview=FitH,colorlinks=true,linkcolor=blue,anchorcolor=black,citecolor=mix,urlcolor=verde]{hyperref}
\usepackage{cite}
\usepackage{color}
\usepackage{xcolor}

\definecolor{verde}{rgb}{0,0.6,0}
\definecolor{mix}{rgb}{1,0.2,0.2}
\newtheorem{property}{Property}

\begin{document}

\begin{titlepage}
     \begin{flushleft}
	\large
	Master 2 Physique Th\'eorique et Math\'ematique,\\
	Physique des Particules et Astrophysique\\
	Aix-Marseille Universit\'e \\
	2014-2015
	\vspace*{1.5cm}
    \end{flushleft}
    \begin{center}
	\vspace*{1cm}
	\huge
	\textbf{Noncommutative Field Theory With General Translation Invariant Star Products}
	\vspace{1.5cm}
	\Large

	\textbf{Manolo Rivera}
	\vspace{1.5cm}

	\includegraphics[width=0.5\textwidth]{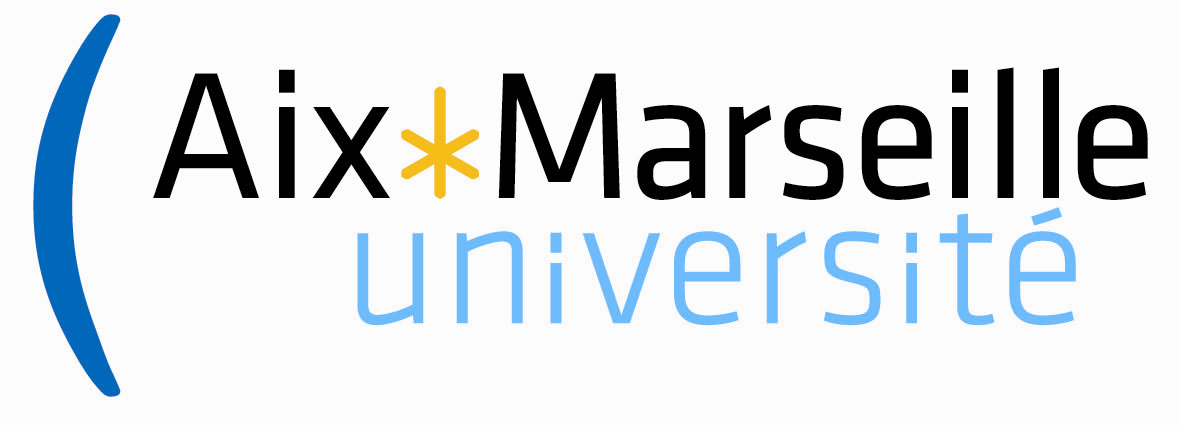}   \includegraphics[width=0.7\textwidth]{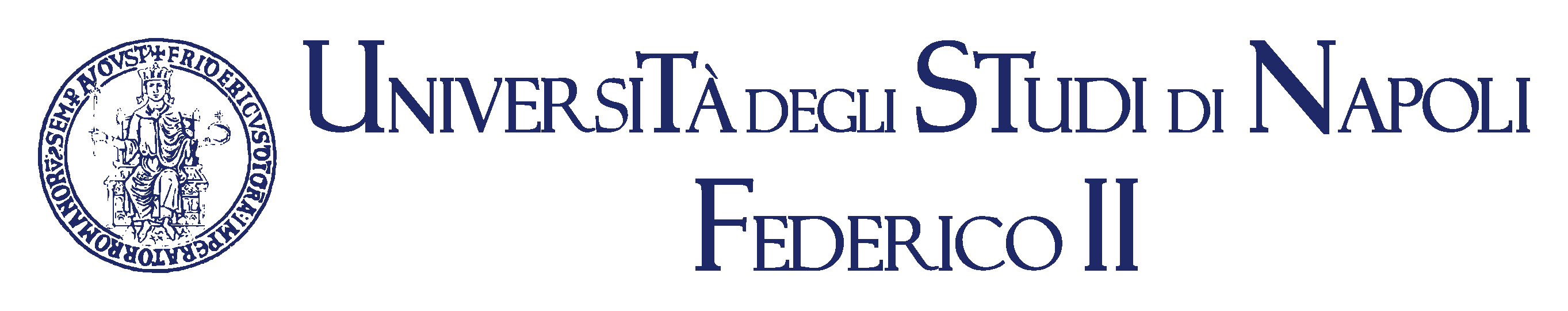}

	\vspace{1.5cm}
	Under the supervision of\\
	\textbf{Prof. Fedele Lizzi} and \textbf{Prof. Patrizia Vitale}\\
	\vspace{0.5cm}
	Universit\`a degli Studi di Napoli Federico II
    \end{center}
\end{titlepage} 

\chapter*{Abstract}

We compute the two-point and four-point Green's function of the noncommutative $\phi^{4}$ field theory; first with the s-ordered star products and then with a general translation invariant star product. We derive the differential expression for any translation invariant star product, and with the help of this expression we show that any of these products can be written in terms of a twist. Finally, using the notion of the twisted action of the infinitesimal Poincar\'e transformations, we show that the commutator between the coordinate functions is invariant under Poincar\'e transformations at a deformed level.    

{\hypersetup{linkcolor=black}
\tableofcontents}

\chapter*{Introduction}\addcontentsline{toc}{chapter}{Introduction}
The idea of studying noncommutative spaces goes back to the fathers of quantum mechanics. It was Erwin Schr\"odinger \cite{refschro} the first one who considered the possibility that geometry looses its meaning in quantum mechanics, but it was Werner Heisenberg \cite{refheis} who suggested the idea of coordinate uncertainty relations to solve the problem of short distance singularities in quantum field theory. The first one who formalized this some years later was Hartland Snyder \cite{snyder}. This idea has been widely studied in the last decades and there are several motivations for that. One of them is that there are strong reasons that suggest that at very small distances (of the order of Planck length), the concept of localization doesn't make sense anymore, i.e. the concept of point is meaningless. Some other motivations for studying noncommutative spaces have come up in other places like string theory \cite{witten} and  constructive field theories \cite{rivasseau}, among others.

The simplest idea of a noncommutative space (or space-time) is to consider a noncommutative $\star$ product \cite{generalstar1, generalstar2, star} (usually called star product) for which the commutator of the coordinate functions is a non-zero constant
	\begin{equation}
	[x^i,x^j]_{\star} =x^i\star x^j-x^j\star x^i= i \theta^{ij}. \label{introeq}
	\end{equation}
The most common star products in the literature are the Gr\"onewold-Moyal \cite{gron,moyal} and the Wick-Voros \cite{voros1,voros2,voros3,voros4}. As we will see later, the Wick-Voros product is associated to the normal ordering while the Moyal product is associated to the symmetric ordering of creation and annihilation operators.  These two are not the only star products that give rise to the commutator (\ref{introeq}). There is a generalization of these two products which is a one-parameter family of star products, called s-ordered products, which range from anti-normal to normal ordering, passing through the symmetric ordering. The s-ordered products haven't been studied as much as the Moyal and Wick-Voros products and this is the reason we will consider them in this work. More precisely, we will consider the $\phi^4$ field theory with s-ordered products and we will compute the two-point and four-point Green's functions up to one loop in detail. For a general review of field theory on noncommutative spaces see \cite{Szabo}.

As we will see, the s-ordered products do not exhaust the star products that give rise to the commutator (\ref{introeq}). The whole family of star products is very wide, however, the relevant star products to be considered in the context of field theory are the translation invariant star products (which include the s-ordered products). These products have been studied in \cite {gallucciolizzivitalemixing, TanasaVitale, ArdalanSadoogi, Varshovi1, LizziVitalereg, Varshovi2}, and it has been shown that the whole family of translation invariant star products can be characterized by a simple expression \cite{gallucciolizzivitalemixing,ArdalanSadoogi}. In this work we realized that most of the computations that can be done for the s-ordered product can be easily generalized to the general translation invariant star products.      

In chapter \ref{chap1}, we briefly introduce the well known Moyal and Wick-Voros products in the context of Weyl and weighted Weyl maps, and we explain how can these two products be generalized to the s-ordered product. For simplicity, we consider the products in a two dimensional space, whose generalization to higher dimensions is straightforward. In chapter \ref{chap-s-ord} we look at classical and quantum field theory with s-ordered products. We consider a $(2+1)$-dimensional space-time where only the two spacial coordinates are non-commutative. We compute the vertex and the propagator of the $\phi^4$ theory and with this we find the two-point and four-point Green's functions up to one loop order and show all the corresponding diagrams. In chapter \ref{chap-transinv} we introduce the general translation invariant star products. We find that all the diagrams and their symmetry factors are the same as for the s-ordered products, and compute explicit expressions of the two-point and four-point green's functions. We also find that the ultraviolet/infrared mixing \cite{mixing1,mixing2,mixing3,mixing4} is present in any noncommutative translation invariant star product, in agreement with \cite{gallucciolizzivitalemixing}. Finally, in chapter \ref{chaptransinv} we will introduce the concept of twist. We show that any translation invariant star product can be written in terms of a twist and show that the star commutator is invariant under the twisted Poincar\'e transformations.

\chapter{Star products}\label{chap1}
In this chapter we will introduce the Moyal and Wick-Voros products, as well as the more general s-ordered product on which we will focus in this and the next chapter. We will first introduce the integral form of the products, and then give the differential form, which is the one we will use for noncommutative field theory. For the moment only the two spacial coordinates are relevant because we consider that the time coordinate commutes with the two spacial coordinates. Time won't be relevant until next chapter where we will do field theory.

\section{Moyal product}

Given two operators $\hat{x}^1$ and $\hat{x}^2$, which satisfy the commutation relation
	\begin{equation}
	[\hat{x}^i,\hat{x}^j] = i \theta^{ij} \label{3}
	\end{equation}
where
	\begin{equation*}
	 \theta^{ij} = (\theta^{-1})_{ij} = \theta \varepsilon^{ij} \qquad \text{with} \qquad \varepsilon^{ij}=  \left(
		\begin{matrix}
		0 & 1 \\
		-1 & 0 
		\end{matrix} \right)
	\end{equation*}
we would like to have a product of functions $f(x^1,x^2) \star g(x^1,x^2)$ such that, in particular, the functions $x^i$ and $x^j$ satisfy the commutation relation (\ref{3}), i,e. $[x^i,x^j]_\star = x^i \star x^j - x^j \star x^i =  i \theta^{ij}$. The Moyal product is among the family of products that satisfy this property, but before defining it, we need to introduce the Weyl map which associates an operator to a function and is defined as 
	\begin{equation}
	\hat{\Omega}_M (f) =\frac{1}{2\pi\theta} \int d^2\alpha\, \tilde{f}(\alpha) W(\alpha) \label{12}
	\end{equation}
where
	\begin{equation}
	W(\alpha) = e^{i\theta_{ij} \hat{x}^i \alpha^j}, \label{13}
	\end{equation}
and
	\begin{equation}
	\tilde{f}(\alpha) =\frac{1}{2\pi\theta} \int d^2 x\, f(x) e^{-i\theta_{ij} x^i \alpha^j } \label{14} 
	\end{equation}
is the symplectic Fourier transform of $f$. So the map can be explicitly written as 
	\begin{equation}
	\hat{\Omega}_{M}(f) =\frac{1}{(2\pi\theta)^2} \int d^2x \, d^2\alpha\, f(x) e^{-i\theta_{ij} x^i \alpha^j } W(\alpha). \label{1}
	\end{equation}
The Weyl map is invertible, and its inverse is given by the map 
	\begin{equation*}
	\Omega_M^{-1}(\hat{A})= \frac{1}{2\pi\theta}\int d^2\beta \, e^{i\theta_{ij} x^i \beta^j } \text{Tr} \left( \hat{A}W^\dag (\beta) \right)
	\end{equation*}
which is called the Wigner map. Indeed,
	\begin{align*}
	\Omega_M^{-1}(\hat{\Omega}_{M}(f) )&= \frac{1}{(2\pi\theta)^2}\int d^2\alpha d^2\beta \, \tilde{f}(\alpha) e^{i\theta_{ij} x^i \beta^j } \text{Tr} \left(W(\alpha) W^\dag (\beta)\right)\\
	&=\, \frac{1}{(2\pi\theta)^2}\int d^2\alpha d^2\beta \, \tilde{f}(\alpha) e^{i\theta_{ij} x^i \beta^j } \int d^2x \, e^{i\theta_{ij} \hat{x}^i (\alpha^j-  \beta^j)} \\
	&= \, \frac{1}{(2\pi\theta)^2}\int d^2\alpha d^2\beta \, \tilde{f}(\alpha) e^{i\theta_{ij} x^i \beta^j }  2\pi  \delta \left(\frac{ \alpha^1-  \beta^1}{\theta} \right) 2\pi \delta \left(\frac{ \alpha^2-  \beta^2}{\theta} \right)\\
	&= \int d^2\alpha d^2\beta \, \tilde{f}(\alpha) e^{i\theta_{ij} x^i \beta^j }  \delta^{(2)} ( \alpha-  \beta) = f(x).
	\end{align*}
This allows to define the Moyal product as
	\begin{equation}
	\hat{\Omega}_M(f\star_M g)= \hat{\Omega}_M(f) \hat{\Omega}_M(g). \label{2}
	\end{equation}
In order to find an integral expression of the Moyal product, note that applying equation (\ref{1}) on the left hand side of the definition, we find  
	\begin{equation}
	\hat{\Omega}_{M}(f\star_M g) =\frac{1}{(2\pi\theta)^2} \int d^2x \, d^2\alpha\, (f\star_M g)(x) e^{-i\theta_{ij} x^i \alpha^j } W(\alpha). \label{4}
	\end{equation}
while the right hand side is
	\begin{align*}
	\hat{\Omega}_{M}(f)\hat{\Omega}_{M}(g) &=\frac{1}{(2\pi\theta)^4} \int d^2y \, d^2\beta\, d^2z \, d^2\gamma\, f(y)g(z) e^{-i\theta_{ij} y^i \beta^j } e^{-i\theta_{ij} z^i \gamma^j } W(\beta)W(\gamma) \\
	=\frac{1}{(2\pi\theta)^4}& \int d^2y \, d^2\beta\, d^2z \, d^2\gamma\, f(y)g(z) e^{-i\theta_{ij} y^i \beta^j } e^{-i\theta_{ij} z^i \gamma^j } e^{\frac{i}{2}\theta_{ij} \beta^i \gamma^j } W(\beta+\gamma). 
	\end{align*}
where we have used the property 
	\begin{equation*}
	W(\beta)W(\gamma) = W(\beta + \gamma) e^{\frac{i}{2}\theta_{ij} \beta^i \gamma^j}
	\end{equation*}
Which is easily found using the Baker-Campbell-Hausdorff formula. Now, using the linear transformations
	\begin{align*}
	\beta &= \alpha -2x+2y \\
	\gamma &= 2x - 2y
	\end{align*}
the product $\hat{\Omega}_{M}(f)\hat{\Omega}_{M}(g) $ takes the form 
	\begin{equation*}
	\frac{4}{(2\pi\theta)^4} \int d^2x \, d^2\alpha\, d^2y \, d^2z\, f(y)g(z) e^{-2i\theta_{ij} (x^i-y^i) (x^j-z^j) } e^{-i\theta_{ij} x^i \alpha^j } W(\alpha). 
	\end{equation*}

Comparing this with equation \eqref{4} we can see that the integral form of the Moyal product is
	\begin{equation}
	(f\star_M g)(x) = \frac{1}{(\pi\theta)^2} \int d^2y \, d^2z\, f(y)g(z) e^{-2i\theta_{ij} (x^i-y^i) (x^j-z^j) }. 
	\end{equation}

\section{Wick-Voros product}

There is a more general version of the Weyl map, called the weighted Weyl map, which is defined as
	\begin{equation*}
	\hat{\Omega}_W(f) =\frac{1}{(2\pi\theta)^2} \int d^2x \, d^2\alpha\, f(x) \omega(\alpha) e^{-i\theta_{ij} x^i \alpha^j } W(\alpha). 
	\end{equation*}
where $\omega(\alpha)$ is an invertible function, called the weighted function. This map is invertible, and its inverse is viven by
	\begin{equation*}
	\Omega_W^{-1}(\hat{A}) =\frac{1}{2\pi\theta} \int d^2\alpha\, \omega^{-1}(\alpha) e^{i\theta_{ij} x^i \alpha^j } \text{Tr} \left( \hat{A}W^{\dag}(\alpha) \right). 
	\end{equation*}

 In the previous section we saw how to define the Moyal product from the Weyl map. The Wick-Voros product is defined in the same way as the Moyal one, but using the following weighted Weyl map
	\begin{equation*}
	\hat{\Omega}_V (f) =\frac{1}{(2\pi\theta)^2} \int d^2x \, d^2\alpha\, f(x) e^{\frac{1}{4\theta}\alpha^2} e^{-i\theta_{ij} x^i \alpha^j } W(\alpha). 
	\end{equation*}
which can be written as

	\begin{equation}
	\hat{\Omega}_V (f) =\frac{1}{(\pi\theta)^2} \int d^2x \, d^2\alpha\, f(x) e^{\frac{1}{2\theta}\alpha_+ \alpha_-} e^{\frac{1}{\theta} (x_+ \alpha_- -\alpha_+ z_-) } W(\alpha).  \label{30}
	\end{equation}
where 

	\begin{equation*}
	x_{\pm} = \frac{x^1 \pm ix^2}{\sqrt{2}} \quad \text{and} \quad \alpha_{\pm} = \frac{\alpha^1 \pm i\alpha^2}{\sqrt{2}}
	\end{equation*}
So, the Wick-Voros product can be defined as
	\begin{equation}
	\Omega_V(f\star_M g)= \hat{\Omega}_V(f) \hat{\Omega}_V(g). 
	\end{equation}
From which the integral form can be shown to be 
	\begin{equation}
	(f\star_V g)(x)= \int \frac{d^2y}{\pi \theta} f(x_-,y_+)g(y_-,x_+)e^{-\frac{1}{\theta} (x_- -y_-) (x_+-z_+) }
	\end{equation}

\section{S-ordered products}
\label{sec:s-ord}

The two operators $\hat{x}^1$ and $\hat{x}^2$ can always be written as
	\begin{equation}
	\hat{x}^1 = \frac{\hat{a}+\hat{a}^\dag}{\sqrt{2}} \quad \text{and} \quad \hat{x}^2 = \frac{\hat{a}-\hat{a}^\dag}{i\sqrt{2}}
	\end{equation}
where $\hat{a}$ and $\hat{a}^\dag$ are two operators which satisfy the commutation relation $[\hat{a},\hat{a}^\dag]=\theta$ and can be seen as the creation and annihilation operators. Using the notation $z=x_+$ and $\omega=\alpha_+$, the Weyl map  \eqref{12} and equations \eqref{13} and \eqref{14} can be written in terms of $z$, $\bar z$, $\omega$, $\bar \omega$, $\hat{a}$ and $\hat{a}^\dag$  as
	\begin{equation}
	\hat{\Omega}_M (f) =\frac{1}{\pi\theta} \int d^2\omega\, \tilde{f}(\omega) W(\omega) ,
	\end{equation}

	\begin{equation}
	W(\omega) =e^{\frac{1}{\theta}(\omega \hat{a}^\dag - \bar{\omega}\hat{a})}
	\end{equation}
and
	\begin{equation}
	\tilde{f}(\omega) =\frac{1}{\pi\theta} \int d^2 z\, f(z,\bar z) e^{\frac{1}{\theta}( z \bar \omega -  \bar{z} \omega)} \label{15}
	\end{equation}

If we write the operators in terms of $\hat{a}$ and $\hat{a}^\dag$, the Weyl map $\hat{\Omega}_M$ gives operators in the symmetric ordering, which means that a monomial $\bar{z}^\alpha z^\beta$ is transformed by $\hat{\Omega}_M$ into the operator
	\begin{equation*}
	\{ (\hat{a}^\dag)^\alpha \hat{a}^\beta \} = \theta^{\alpha+\beta}\frac{\partial^{\alpha+\beta}W(\omega)}{\partial \omega^\alpha \partial (-\bar{\omega})^\beta} \bigg\vert_{\omega=0}.
	\end{equation*}
For example the monomial $\bar{z} z$ is sent to $ \{ \hat{a}^\dag \hat{a} \} = \frac{1}{2}(\hat a \hat a^\dag+\hat a^\dag \hat a) $. On the other hand, the weighted Weyl map $\hat{\Omega}_V$ corresponding to the Wick-Voros product, gives always operators in the normal ordering (annihilators to the right and creators to the left).  A third way of ordering the operators is the antinormal ordering (creators to the right and annihilators to the left). These are particular cases of a more general ordering which is defined as
	\begin{equation*}
	\{ (\hat{a}^\dag)^\alpha \hat{a}^\beta \}_s =  \theta^{\alpha+\beta} \frac{\partial^{\alpha+\beta}W_s(\omega)}{\partial \omega^\alpha \partial (-\bar{\omega})^\beta} \bigg\vert_{\omega=0}
	\end{equation*}
where
	\begin{equation}
	W_s(\omega) =e^{\frac{s }{2\theta}\lvert \omega \lvert^2}W(\omega) \label{21}
	\end{equation}
and $s \in [-1,1]$. In particular, for $s=-1,0 \; \text{and} \, 1$, the monomial $\bar{z} z$ is sent to the antinormal, symmetric and normal ordered operators respectively
	\begin{equation*}
	\{ \hat{a}^\dag \hat{a} \}_{-1} =  \hat{a} \hat{a}^\dag, \quad \{ \hat{a}^\dag \hat{a} \}_0 = \frac{1}{2}(\hat a \hat a^\dag+\hat a^\dag \hat a) \quad \text{and} \quad \{ \hat{a}^\dag \hat{a} \}_1 =   \hat{a}^\dag \hat{a}.
	\end{equation*} 

There is a one parameter family of weighted Weyl maps that give rise to s-ordered operators, and is given by
	\begin{equation}
	\hat{\Omega}_s (f) =\frac{1}{\pi\theta} \int d^2\omega\, \tilde{f}(\omega) W_s(\omega) \label{16}
	\end{equation}
with $s \in [-1,1]$. Note that this map reduces to the Weyl map \eqref{12} for $s=0$, and to the weighted Weyl map corresponding to the Wick-Voros product \eqref{30} for $s=1$.  This map can be explicitly written as 
	\begin{equation*}
	\hat{\Omega}_s (f) =\frac{1}{(\pi\theta)^2} \int d^2\omega\, d^2 z\,  f(z,\bar z) e^{\frac{1}{\theta}(z \bar \omega - \bar{z} \omega)}  W_s(\omega)
	\end{equation*}
or equivalently
	\begin{equation}
	\hat{\Omega}_s (f) =\frac{1}{\pi\theta} \int d^2z\,  f(z,\bar z) \breve{W}_s(z)
	\end{equation}
where
	\begin{equation}
	\breve{W}_s(z) = \frac{1}{\pi\theta} \int d^2 \omega\,  e^{\frac{1}{\theta}(z \bar \omega - \bar{z} \omega)}  W_s(\omega)
	\end{equation}  
is another kind of symplectic Fourier transform, like \eqref{15}. This map is invertible, and its inverse is given by 
	\begin{equation}
	\left( \Omega_s^{-1} (\hat A) \right) (z) = \text{Tr} \left( \hat{A} \breve{ W}_{-s}(z) \right) \label{20}
	\end{equation}
To prove this we need the following properties (see appendix \ref{sec:appB})
	\begin{equation}
	W(\omega) W(\omega ') = e^{\frac{i}{\theta} \text{Im}(\omega \bar {\omega}')}W(\omega + \omega ') \label{17}
	\end{equation}
and
	\begin{equation}
	\text{Tr} \, W(\omega) = \pi\theta \delta(\omega) \label{19}
	\end{equation}
We can now check equation \eqref{20}
	\begin{align*}
	\left( \Omega_s^{-1} (\hat{\Omega}_s (f)) \right) (z') &= \text{Tr} \left( \hat{\Omega}_s (f) \breve{ W}_{-s}(z') \right) \\
	&= \frac{1}{\pi\theta} \int d^2z\,  f(z,\bar z)  \text{Tr} \left( \breve{W}_s(z)   \breve{W}_{-s}(z') \right)\\
	= \frac{1}{(\pi\theta)^3} \int d^2z\,& d^2 \omega\, d^2 \omega'\,   f(z,\bar z) e^{\frac{s }{2\theta}(\lvert \omega \lvert^2-\lvert \omega' \lvert^2)}  e^{\frac{1}{\theta}(z \bar \omega - \bar{z} \omega)} e^{\frac{1}{\theta}(z' \bar \omega' - \bar{z'} \omega')} \text{Tr} \left(  W(\omega) W(\omega') \right)
	\end{align*}
but using properties \eqref{17} and \eqref{19} this is
	\begin{align*}
	&= \int \frac{ d^2z\, d^2 \omega\, d^2 \omega'}{(\pi\theta)^3}   f(z,\bar z) e^{\frac{s }{2\theta}(\lvert \omega \lvert^2-\lvert \omega' \lvert^2)}  e^{\frac{1}{\theta}(z \bar \omega - \bar{z} \omega+z' \bar \omega' - \bar{z'} \omega)} e^{\frac{i}{\theta} \text{Im}(\omega \bar {\omega}')} \text{Tr} \left( W(\omega + \omega ') \right) \\
	&= \int \frac{ d^2z\, d^2 \omega\, d^2 \omega'}{(\pi\theta)^3}   f(z,\bar z) e^{\frac{s }{2\theta}(\lvert \omega \lvert^2-\lvert \omega' \lvert^2)}  e^{\frac{1}{\theta}(z \bar \omega - \bar{z} \omega+z' \bar \omega' - \bar{z'} \omega)} e^{\frac{i}{\theta} \text{Im}(\omega \bar {\omega}')} \pi\theta \delta(\omega + \omega ') \\
	&= \int \frac{ d^2z\, d^2 \omega}{(\pi\theta)^2}   f(z,\bar z)   e^{\frac{1}{\theta}((z-z') \bar \omega - (\bar{z}-\bar{z'}) \omega)} = \int \frac{ d^2z}{(\pi\theta)^2}   f(z,\bar z)   \pi^2 \delta^{(2)} \left(\frac{z-z'}{\theta}\right)  = f(z',\bar z').
	\end{align*}

Given that the map $\hat{\Omega}_s$ is invertible, we can define the s-ordered star product as
	\begin{equation}
	(f \star_s g)(z,\bar z) =  \text{Tr} \left( \hat\Omega (f) \hat\Omega(g) \breve{W}_{-s}(z) \right) . \label{100}
	\end{equation}
It has been shown \cite{soloviev} that this product can be written as a series expansion as follows
	\begin{equation}
	(f \star_s g)(z,\bar z) =  f(z,\bar z)   e^{\frac{\theta}{2}((s+1)\overleftarrow\partial_{z} \overrightarrow\partial_{\bar z}+(s-1)\overleftarrow\partial_{\bar z}. \overrightarrow\partial_{z})} g(z,\bar z). \label{23}
	\end{equation}
This is what we call the differential expression of the s-ordered product. It is important to mention that the range of the differential expression \cite{soloviev} is smaller than the range of the integral expression. From now on we assume that the functions under consideration are in the range of the differential expression.

\chapter{Field Theory with s-ordered star products} \label{chap-s-ord}

In this chapter we study the noncommutative $\phi^4$ field theory obtained from the commutative one by replacing the ordinary product with the s-ordered star product. In particular we consider the following action
	\begin{equation}
	S = S_0 - S_{\text{int}}  \label {7}
	\end{equation}
where $S_0$ is the free Klein-Gordon action given by 
	\begin{equation}
	S_0 = \int d^3x \frac{1}{2}(\partial_\mu \phi \partial^\mu \phi - m^2 \phi^2)
	\end{equation}
and $S_{int}$ is the interacting action given by
	\begin{equation}
	S_{\text{int}} =  \frac{g}{4!} \int d^3x \; \phi^4.
	\end{equation}

Now, as we said, we consider the noncommutative action by replacing the ordinary product by the star product. So the free action becomes
	\begin{equation}
	S_{0_s} = \int d^3x \frac{1}{2}(\partial_\mu \phi  \star_s \partial^\mu \phi - m^2 \phi  \star_s \phi)
	\end{equation}
while the interacting action becomes
	\begin{equation}
	S_{\text{int}_s} =  \frac{g}{4!} \int d^3x \; \phi  \star_s \phi  \star_s \phi  \star_s \phi. \label{8}
	\end{equation}
We will compute the two and four point Green's functions of this theory, but before going to the quantum level, let us look at some properties of the theory at the classical level.

\section{Classical Field Theory} \label{classfieldt}

In this section we will describe some properties of the theory at the classical level that will be useful for the next section, where we will compute the propagator and the vertex of the theory at quantum level. We start by looking at the following property
  	\begin{equation*}
	\int d^3 x \, f  \star_s g= \int d^3 x \; g  \star_s f
	\end{equation*}  
which is called the trace property. To check this we will use the differential form of the star product \eqref{23}, which written in terms of $x_+$ and $x_-$ is
	\begin{equation}
	(f \star_s g)(x) =  f(x)   e^{\frac{\theta}{2}((s+1)\overleftarrow\partial_{x^+} \overrightarrow\partial_{x^-}+(s-1)\overleftarrow\partial_{x^-} \overrightarrow\partial_{x^+})} g(x), \label{18}
	\end{equation}
so we can write the star product as
  	\begin{align}
	 f  \star_s g&= \int \frac{d^3 p}{(2\pi)^3} \frac{d^3 q}{(2\pi)^3} \;  \tilde{f}(p) \tilde{g}(q) e^{-i p\cdot x} \star_s e^{-i q\cdot x}\nonumber \\
	 &= \int \frac{d^3 p}{(2\pi)^3} \frac{d^3 q}{(2\pi)^3} \;  \tilde{f}(p) \tilde{g}(q) e^{-i (p_+ x_- + p_- x_+)} \star_s e^{-i (q_+ x_- + q_- x_+)}\nonumber \\
	&= \int \frac{d^3 p}{(2\pi)^3} \frac{d^3 q}{(2\pi)^3} \;  \tilde{f}(p) \tilde{g}(q) e^{-\frac{\theta}{2} [(s+1)p_- q_+ + (s-1) p_+ q_-]}  e^{-i (p+q)\cdot x}.  \label{9}
	\end{align}  
Integrating over $x$ we get
  	\begin{equation}
	 \int d^3x \, f  \star_s g = \int \frac{d^3 p}{(2\pi)^3} \;  \tilde{f}(p) \tilde{g}(-p) e^{\theta s \, p_- p_+ }
	\end{equation}  
so the star product commutes inside the integral, which is what we wanted to prove. The other property we will need is the explicit expression of the star product inside the integral. For this we use again the differential form of the s-ordered product \eqref{18}
  	\begin{equation*}
	\int d^3 x (f  \star_s g)(x) = \int d^3 x f(x) e^{\frac{\theta}{2} [(s+1) \overleftarrow{\partial}_{x^+} \overrightarrow{\partial}_{x^-}+(s-1) \overleftarrow{\partial}_{x^-} \overrightarrow{\partial}_{x^+}]}  g(x)
	\end{equation*}  
	\begin{equation*}
	 = \int d^3 x f(x) \sum_{n=0}^\infty \frac{\left( \frac{\theta}{2} [(s+1) \overleftarrow{\partial}_{x^+} \overrightarrow{\partial}_{x^-}+(s-1) \overleftarrow{\partial}_{x^-} \overrightarrow{\partial}_{x^+}] \right)^n}{n!} g(x)
	\end{equation*}
  	\begin{equation*}
	= \int d^3 x f(x) \sum_{n=0}^\infty  \frac{(\theta / 2)^n}{n!} \left( \sum_{i=0}^n \dbinom{n}{i} (s+1)^i \overleftarrow{\partial}_{x^+}^i \overrightarrow{\partial}_{x^-}^i (s-1)^{n-i} \overleftarrow{\partial}_{x^-}^{n-i} \overrightarrow{\partial}_{x^+}^{n-i} \right) g(x)
	\end{equation*}
  	\begin{equation*}
	= \int d^3 x \sum_{n=0}^\infty  \frac{(\theta / 2)^n}{n!} \left( \sum_{i=0}^n \dbinom{n}{i} (s+1)^i  (s-1)^{n-i} (\partial_{x^+}^i \partial_{x^-}^{n-i} f(x)) (\partial_{x^-}^i  \partial_{x^+}^{n-i} g(x)) \right) 
	\end{equation*}
so, doing integration by parts and neglecting the boundary terms we have
  	\begin{equation*}
	= \int d^3 x \sum_{n=0}^\infty  \frac{(\theta / 2)^n}{n!} \left( \sum_{i=0}^n \dbinom{n}{i} (s+1)^i  (s-1)^{n-i}  (-1)^n f(x) \partial_{x^-}^n  \partial_{x^+}^n g(x) \right) 
	\end{equation*}
but noting that $\sum_{i=0}^n \dbinom{n}{i} (s+1)^i  (s-1)^{n-i} = (2s)^n$ we get
  	\begin{equation*}
	= \int d^3 x \; f(x) \; \sum_{n=0}^\infty  \frac{(-s\theta )^n}{n!} (\overrightarrow{\partial}_{x^-} \overrightarrow{\partial}_{x^+})^n g(x)  = \int d^3 x \, f(x) \,  e^{ -s\theta  (\overrightarrow{\partial}_{x^-} \overrightarrow{\partial}_{x^+})} g(x)
	\end{equation*}
finally, using $\partial_{x^+}\partial_{x^-}= \frac{1}{2}\nabla^2$, we get the following expression for the integral of the s-ordered star product
   	\begin{equation}
	\int d^3 x (f  \star_s g)(x) = \int d^3 x \, f(x) \,  e^{-\frac{1}{2} s \theta \nabla^2} g(x). \label{5}
	\end{equation}  
We can use this results to find the equation of motion. First we need to make a small variation $\phi \to \phi + \delta \phi $ so that the variation of the action is 
	\begin{equation*}
	\delta S_{0_s} = \int d^3x (\partial_\mu \delta \phi  \star_s \partial^\mu \phi - m^2 \delta \phi  \star_s \phi).
	\end{equation*}
where we used the trace property. Integrating by parts and neglecting the boundary terms this is
	\begin{equation}
	\delta S_{0_s} = - \int d^3x  \delta \phi  \star_s ( \Box + m^2) \phi . \label{38}
	\end{equation}
Using equation \eqref{5} we have
	\begin{equation*}
	\delta S_{0_s} = - \int d^3x \;  \delta \phi \,  e^{-\frac{1}{2} s \theta \nabla^2} ( \Box + m^2) \phi 
	\end{equation*}
but this must vanish for any variation of the field $\delta \phi$, so the equation of motion is
	\begin{equation}
	  e^{-\frac{1}{2} s \theta \nabla^2} ( \Box + m^2) \phi = 0 . \label{6}
	\end{equation}

Note that the only case for which this equation is equal to the Klein-Gordon equation is for $s=0$, however, both equations have the same solutions due to the invertibility of the operator $e^{\frac{1}{2} s \theta \nabla^2}$. Moreover, the on shell condition is the same, to see this let us take the Fourier transform of equation \eqref{6}
	\begin{align*}
	  e^{-\frac{1}{2} s \theta \nabla^2} ( \Box + m^2) \phi &=  e^{-\frac{1}{2} s \theta \nabla^2} ( \Box + m^2) \int \frac{d^3k}{(2 \pi)^3} \tilde{\phi} (k) e^{-ik\cdot x} \\
	&= \int \frac{d^3k}{(2 \pi)^3} e^{\frac{1}{2} s \theta \boldsymbol{k}^2} (- k^2 + m^2)  \tilde{\phi} (k) e^{-ik\cdot x} = 0
	\end{align*}
where we use the notation $k^2 = k^\mu k_\mu$ for $\mu= 0,1,2$ and $\boldsymbol{k}^2 =  k^i k_i$ for $i= 1,2$. From now on, the boldface characters will stand for vectors in the two-dimensional space, while normal characters will stand for vectors in the (2+1)-dimensional space-time. The on shell condition is then given by 
	\begin{equation*}
	 e^{\frac{1}{2} s \theta \boldsymbol{k}^2} ( k^2 - m^2)  \tilde{\phi} (k)= 0
	\end{equation*}
which is equivalent to the usual condition
	\begin{equation}
	  ( k^2 - m^2)  \tilde{\phi} (k)= 0
	\end{equation}
which means that, at the classical level, the noncommutative free field theory given by the action \eqref{7} with the s-ordered star product, is the same as the commutative one.

\section{Quantum Field Theory}

We want to compute the Green's functions of the theory, but for that we need the propagator and the vertex which we proceed to compute in this section.

\subsection{Propagator}
To compute the propagator, which we call $G_s$, we start from its definition
	\begin{equation*}
	  e^{-\frac{1}{2} s \theta \nabla^2} ( \Box + m^2) G_s (x-x') = -\delta^{(3)} (x-x').
	\end{equation*}
Writing it in terms of the Fourier transform, the left hand side becomes
 	\begin{align*}
	  e^{-\frac{1}{2} s \theta \nabla^2} ( \Box + m^2) G_s (x-x') &=  e^{-\frac{1}{2} s \theta \nabla^2} ( \Box + m^2) \int \frac{d^3 p}{(2\pi)^3} \tilde{G}_s (p) e^{-i p \cdot (x-x')}\\
	& =   \int \frac{d^3 p}{(2\pi)^3} e^{\frac{1}{2} s \theta \boldsymbol{p}^2} (- p^2 + m^2) \tilde{G}_s (p) e^{-i p \cdot (x-x')}
	\end{align*}
while the right hand side is
 	\begin{equation*}
	 -\delta^{(3)} (x-x') =  - \int \frac{d^3 p}{(2\pi)^3} e^{-i p \cdot (x-x')}
	\end{equation*}
from which follows that 
 	\begin{equation*}
	  e^{\frac{1}{2} s \theta \boldsymbol{p}^2} (- p^2 + m^2) \tilde{G}_s (p)=-1
	\end{equation*}
so the propagator is
 	\begin{equation}
	  \tilde{G}_s (p)= \frac{ e^{-\frac{1}{2} s \theta \boldsymbol{p}^2}}{ p^2 - m^2}. \label{11}
	\end{equation}

\subsection{The vertex}

In order to find the pour-point Green's function at tree level, we need to compute the vertex in momentum space. For that we need to compute an explicit expression for the star product of exponentials. Using equation \eqref{9} he have
  	\begin{equation*}
	 e^{-i(k_1+k_2)\cdot x} \star_s  e^{-i(k_3+k_4)\cdot x} =
	\end{equation*}
  	\begin{equation*}
	 = \int \frac{d^3 p}{(2\pi)^3} \frac{d^3 q}{(2\pi)^3} \;  (2\pi)^6 \delta^{(3)}(k_1+k_2-p)  \delta^{(3)}(k_3+k_4-q) e^{-\frac{\theta}{2} [(s+1)p_- q_+ + (s-1) p_+ q_-]}  e^{-i (p+q)\cdot x} 
	\end{equation*}
  	\begin{equation*}
	 = e^{-\frac{\theta}{2}[(s+1) (k_{1_-}+k_{2_-})  (k_{3_+}+k_{4_+}) +(s-1) (k_{1_+}+k_{2_+})  (k_{3_-}+k_{4_-}) ]}e^{-i(k_1+k_2+k_3+k_4)\cdot x}.
	\end{equation*}
Using this expression and equation \eqref{9}, we have
	\begin{equation*}
	S_{\text{int}_s} =  \frac{g}{4!} \int d^3x \; (\phi  \star_s \phi)  \star_s (\phi  \star_s \phi) 
	\end{equation*}
	\begin{eqnarray*}
	=  \frac{g}{4!}  \int d^3x \prod_{i=1}^4 \frac{d^3 k_i}{(2\pi)^3} \;  \tilde{\phi}(k_i) \;  e^{-\frac{\theta}{2} [(s+1)k_{1_-} k_{2_+} + (s-1) k_{1_+} k_{2_-}]} e^{-\frac{\theta}{2} [(s+1)k_{3_-} k_{4_+} + (s-1) k_{3_+} k_{4_-}]} \\  \left( e^{-i(k_1+k_2)\cdot x} \star_s  e^{-i(k_3+k_4)\cdot x} \right)
	\end{eqnarray*}
	\begin{eqnarray*}
	=  \frac{g}{4!}  \int d^3x \prod_{i=1}^4 \frac{d^3 k_i}{(2\pi)^3} \;  \tilde{\phi}(k_i) \;  e^{-\frac{\theta}{2} [(s+1)k_{1_-} k_{2_+} + (s-1) k_{1_+} k_{2_-}]} e^{-\frac{\theta}{2} [(s+1)k_{3_-} k_{4_+} + (s-1) k_{3_+} k_{4_-}]} \\   e^{-\frac{\theta}{2}[(s+1) (k_{1_-}+k_{2_-})  (k_{3_+}+k_{4_+}) +(s-1) (k_{1_+}+k_{2_+})  (k_{3_-}+k_{4_-}) ]}e^{-i(k_1+k_2+k_3+k_4)\cdot x}
	\end{eqnarray*}
	\begin{equation*}
	=  \frac{g(2\pi)^3}{4!}  \int \prod_{i=1}^4 \frac{d^3 k_i}{(2\pi)^3} \;  \tilde{\phi}(k_i) \;  e^{-\frac{\theta}{2} \sum_{a<b} [(s+1) k_{a_-} k_{b_+} + (s-1) k_{a_+} k_{b_-}]} \delta^{(3)}(k_1+k_2+k_3+k_4).
	\end{equation*}
Finally we have
	\begin{equation}
	S_{\text{int}_s}= i \int  \prod_{i=1}^4 \frac{d^3 k_i}{(2\pi)^3} \;  \tilde{\phi}(k_i) \;  V_{\star_s} \label{31}
	\end{equation}
where 
	\begin{equation}
	 V_{\star_s} = V  e^{-\frac{\theta}{2} \sum_{a<b} [(s+1) k_{a_-} k_{b_+} + (s-1) k_{a_+} k_{b_-}]} \label{10}
	\end{equation}
is the vertex of the theory, and 

	\begin{equation}
	 V = -i \frac{g(2\pi)^3}{4!}  \delta^{(3)}(k_1+k_2+k_3+k_4) \label{101}
	\end{equation}
is the usual vertex of the commutative theory. Note that the vertex is not invariant under an arbitrary permutation of the momenta, however, it is invariant under cyclic permutations\footnote{Here by cyclic permutation we mean $\sigma \in S_n$ of the form $\sigma(a) = (a + k)mod\, n$} (see appendix \ref{sec:app}). It is important to compare with the Moyal case. First note that defining 
	\begin{equation}
	 p \wedge q = \varepsilon^{ij} p_i q_j
	\end{equation}
the vertex can be written as
	\begin{equation}
	 V_{\star_s} = V  e^{-\frac{\theta}{2} \sum_{a<b} [s \boldsymbol k_a \cdot \boldsymbol k_b+i \boldsymbol k_a \wedge \boldsymbol k_b]}  \label{33}
	\end{equation}
where we used the property $p_- q_+ = \frac{1}{2} (\boldsymbol p\cdot \boldsymbol q + i \boldsymbol p \wedge \boldsymbol q)$.  The Moyal vertex is given by the same expression with $s=0$
 	\begin{equation}
	 V_{\star_M}= V_{\star_0} = V  e^{-\frac{\theta}{2} \sum_{a<b} i \boldsymbol k_a \wedge \boldsymbol k_b}. \label{34}
	\end{equation}

Using conservation of momentum $\sum k_a =0$ it is easy to show that $ \sum_{a<b}  \boldsymbol k_a \cdot \boldsymbol k_b=-\frac{1}{2}\sum \boldsymbol k^2_a$, which is invariant under any permutation. So the Moyal part determines the symmetries of the vertex, the non-Moyal part is invariant under any permutation. 

\section{Green's functions}

We have the vertex and the propagator, therefore we can compute the two-point and four-point Green's functions. At zeroth order we have the same diagrams as in conventional quantum field theory (QFT) except that the propagators are the ones we found in noncommutative quantum field theory (NCQFT). At first order things get a bit different because we already have a vertex (which as we saw, is not invariant under a generic permutation), so we need to keep track of the order in which the lines enter the vertex. This is why we will have to consider some diagrams that are equivalent in ordinary QFT, but are not equivalent in NCQFT \cite{IR/UVmixing}. In order to see this, and to compute the symmetry factors of the diagrams, we need to recall how Green's functions in momentum space are computed from the generating function (see \cite{peskin}, \cite{lebellac})\footnote{Note that unlike \cite{lebellac}, we include the delta inside the Green's function}  
 	\begin{equation}
	\begin{split}
	(2\pi)^3 G^{(N)}(k_1,...,k_N) =(2\pi)^{3N/2} \text{exp} \left( -iS_{\text{int}_s} \left(\frac{\delta}{i\,\delta j(-q)} \right) \right)  \times \qquad \qquad \qquad \qquad \\  \frac{\delta^N}{i\,\delta j(-k_1) \dots i\,\delta j(-k_N)} \; \text{exp} \left(- \frac{1}{2} \int d^3 k \; j(k) G_s (k) j(-k) \right) \label{32}
	\end{split}
	\end{equation}
where 
 	\begin{equation*}
	S_{\text{int}_s} \left(\frac{\delta}{i\,\delta j(-q)} \right) = i \int \left[ \prod_{i=1}^4 \frac{d^3 q_i}{(2\pi)^{3/2}} \frac{\delta}{\delta j(-q_i)} \right]  V_{\star_s}.
	\end{equation*}
The computation of the Green's functions is done in the same way as in QFT, but as we said we have to keep track of the order of the lines in each vertex due to the factor $\text{exp}(-\frac{\theta}{2} \sum_{a<b} [(s+1) k_{a_-} k_{b_+} + (s-1) k_{a_+} k_{b_-}])$ which comes inside $V_{\star_s}$. 

Here we are interested in the connected component of the Green's function $G_c$, which is the relevant part if we want to compute scattering amplitudes.

\subsection{Two-point Green's function} \label{2pointgf}

In order to see how all this works, we will do the computation of the two-point correlation function $\tilde G_c^{(2)}$ in detail. From equation \eqref{32} we can see that $\tilde G_c^{(2)}(p)$ (where $p=k_1=-k_2$) at leading order is
 	\begin{equation*}
	\tilde G_{c; \, 0}^{(2)}(p) = \tilde G_s (p) = \frac{ e^{-\frac{1}{2} s \theta \boldsymbol{p}^2}}{ p^2 - m^2}
	\end{equation*}
where the second subindex refers to the order of expansion. So, as usual it is just the propagator, as we expected because there is no vertex. At one loop order, things get a bit more complicated than in the commutative case because of the vertex.

\begin{figure}[htb]
	\begin{subfigure}{0.5\textwidth}
		\includegraphics[scale=0.45]{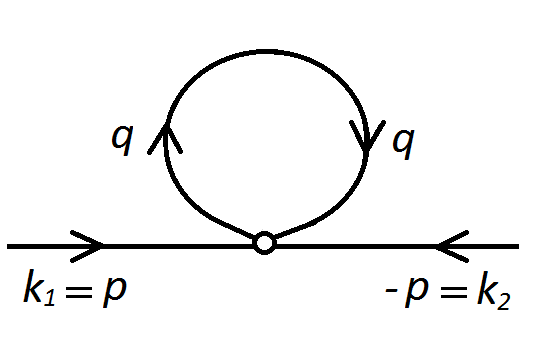}
		\caption{} \label{fig1a}
	\end{subfigure}
\hspace*{1cm}
	\begin{subfigure}{0.5\textwidth}
		\includegraphics[scale=0.45]{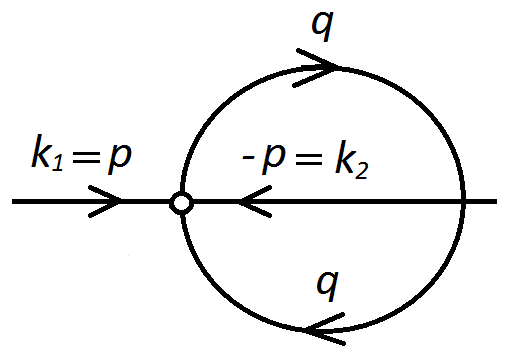}
		\caption{} \label{fig1b}
	\end{subfigure}
	\caption{\sl Two-point Green's function one loop diagrams} \label{fig1}
\end{figure}

\begin{figure}[htb]

	\begin{subfigure}{0.3\textwidth}
		\hspace*{1cm}
		\includegraphics[scale=0.4]{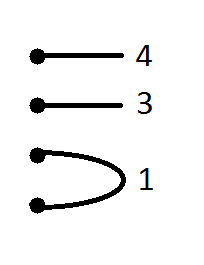}
		\caption{} \label{fig2a}
	\end{subfigure}
	\begin{subfigure}{0.3\textwidth}
		\hspace*{1cm}
		\includegraphics[scale=0.4]{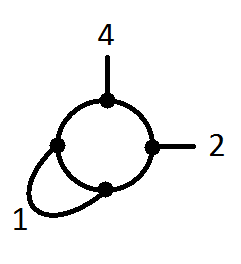}
		\caption{} \label{fig2b}
	\end{subfigure}
	\begin{subfigure}{0.3\textwidth}
		\hspace*{1cm}
		\includegraphics[scale=0.4]{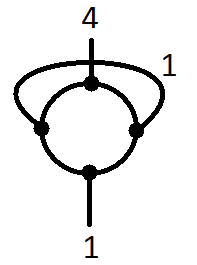}
		\caption{} \label{fig2c}
	\end{subfigure}

	\caption{\sl Two-point Green's function vertex. Figure (a) corresponds to the vertex in QFT, figures (b) and (c) correspond to the two inequivalent vertices in NCQFT. The numbers represent the number of ways of attaching the corresponding line to the vertex} \label{fig2}
\end{figure}

 In QFT we would just have the diagram shown in figure \ref{fig1a}
with a symmetry factor $12/4!$. The usual way to compute this symmetry factor is by thinking of the vertex as made by four points, and then count the number of ways to attach the four lines to the vertex, as it is shown in figure \ref{fig2a}
. In NCQFT we do the same but we just count the number of ways of attaching the lines in such a way that the vertex doesn't change. Due to the invariance under cyclic permutations, just the order of the lines is relevant, so we can think of the four points as being on a ring, like in figures \ref{fig2b} and \ref{fig2c}.  Saying that the vertex is invariant under cyclic permutations is like saying that rotating the ring doesn't change anything. Let us use this to compute  $\tilde G_c^{(2)}$. Using the fact that $K_1=-K_2$ it is easy to see that there are just two inequivalent vertices, shown in figures \ref{fig2b} and \ref{fig2c}, and their corresponding diagrams are shown in figures \ref{fig1a} and \ref{fig1b} respectively. The symmetry factor of \ref{fig2b} is $8/4!$ because there are four ways of attaching the first external line, two ways of attaching the second external line (to the right or to the left of the first one) and the internal loop joins the two remaining points. On the other hand, the symmetry factor of \ref{fig2c} is $4/4!$ because there are four ways for the first external line, the second one must be in the opposite point and the internal loop joins the two remaining points. Of course the sum of the two symmetry factors is $12/4!$ which is the symmetry factor in QFT.    

The integral corresponding to a given diagram is found in the same way as in ordinary QFT (as can be seen from equation \eqref{32}), i.e. the Feynman rules are the same except that the symmetry factors are computed in the way we explained, and the vertex and propagator are given by equations \eqref{33} and \eqref{11}. For the diagram \ref{fig1a}, the vertex is given by\footnote{To compute the phase of the vertex we choose any line (due to cyclic invariance) and then move counterclockwise} 
	\begin{equation*}
	 V_{\star_s} = V e^{-\frac{\theta}{2} [s (-\boldsymbol q \cdot \boldsymbol q +\boldsymbol q\cdot \boldsymbol p -\boldsymbol q\cdot \boldsymbol p -\boldsymbol q\cdot \boldsymbol p +\boldsymbol q\cdot \boldsymbol p -\boldsymbol p \cdot \boldsymbol p)+i (-\boldsymbol q \wedge \boldsymbol q +\boldsymbol q\wedge \boldsymbol p -\boldsymbol q\wedge \boldsymbol p -\boldsymbol q\wedge \boldsymbol p +\boldsymbol q\wedge \boldsymbol p -\boldsymbol p \wedge \boldsymbol p)]} 
	\end{equation*}
	\begin{equation*}
	 = V  e^{\frac{\theta s}{2}  (\boldsymbol q^2+\boldsymbol p^2)}.
	\end{equation*}
The integral corresponding to the diagram \ref{fig1a} is
	\begin{align}
	 \tilde{G}_{\begin{NoHyper}\ref{fig1a}\end{NoHyper}}^{(2)}(p)&= -ig\frac{8}{4!} \int \frac{d^3 q}{(2\pi)^3}\, \frac{ e^{-\frac{\theta s}{2} ( 2\boldsymbol{p}^2+ \boldsymbol{q}^2)} e^{\frac{\theta s}{2} ( \boldsymbol{q}^2+ \boldsymbol{p}^2)}}{ (p^2 - m^2)^2 (q^2 - m^2)} \nonumber \\ &= \frac{-ig}{3} \int \frac{d^3 q}{(2\pi)^3}\, \frac{ e^{-\frac{\theta s}{2}  \boldsymbol{p}^2}}{ (p^2 - m^2)^2 (q^2 - m^2)}. \label{50}
	\end{align}
Doing the same for the diagram  \ref{fig1b} we find
	\begin{equation}
	 \tilde{G}_{\begin{NoHyper}\ref{fig1b}\end{NoHyper}}^{(2)}(p)=  \frac{-ig}{6} \int \frac{d^3 q}{(2\pi)^3}\, \frac{ e^{-\theta (\frac{s}{2} \boldsymbol{p}^2 + i \boldsymbol p\wedge \boldsymbol q)}}{  (p^2 - m^2)^2  (q^2 - m^2)}. \label{51}
	\end{equation}
The connected two-point Green's function up to one loop order is the propagator plus the sum of these two integrals with their corresponding symmetry factors
 	\begin{align}
	  \tilde{G}_{c; \, 1}^{(2)}(p)&= \frac{ e^{-\frac{1}{2} s \theta \boldsymbol{p}^2}}{ p^2 - m^2} - \frac{ig}{4!} \int \frac{d^3 q}{(2\pi)^3}\, \frac{ 8\, e^{-\frac{\theta s}{2} \boldsymbol{p}^2} +  4\, e^{- \theta (\frac{s}{2} \boldsymbol{p}^2 + i \boldsymbol p\wedge \boldsymbol q)} }{  (p^2 - m^2)^2  (q^2 - m^2)} \nonumber\\
	 &=\frac{ e^{-\frac{1}{2} s \theta \boldsymbol{p}^2}}{ p^2 - m^2} - \frac{ig}{6} \int \frac{d^3 q}{(2\pi)^3}\, \frac{( 2\, +  e^{- \theta i \boldsymbol p\wedge \boldsymbol q})e^{-\frac{1}{2} s \theta \boldsymbol{p}^2} }{  (p^2 - m^2)^2 (q^2 - m^2)}.  \label{45}
	\end{align}
Note that if we set $\theta =0$ we get the conventional expression 
 	\begin{equation*}
	 \tilde{G}_1^{(2)}(p)=\frac{ 1}{ p^2 - m^2} - \frac{ig}{2} \int \frac{d^3 q}{(2\pi)^3}\, \frac{1 }{  (p^2 - m^2)^2 (q^2 - m^2)} .
	\end{equation*}

\subsection{Four-point Green's function} \label{4pointgf}

\begin{figure}[htb]
\renewcommand*\thesubfigure{\arabic{subfigure}}
\makeatletter
\renewcommand{\p@subfigure}{\thefigure-}
\makeatother
\hspace*{0.5cm}
	\begin{subfigure}{0.3\textwidth}
\hspace*{0.5cm}
		\includegraphics[scale=0.4]{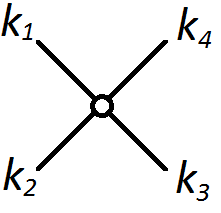}
		\caption{} \label{fig3a}
	\end{subfigure}
	\begin{subfigure}{0.3\textwidth}
\hspace*{0.5cm}
		\includegraphics[scale=0.4]{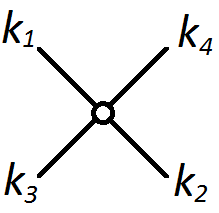}
		\caption{} \label{fig3b}
	\end{subfigure}
	\begin{subfigure}{0.3\textwidth}
\hspace*{0.5cm}
		\includegraphics[scale=0.4]{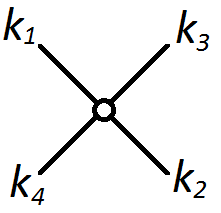}
		\caption{} \label{fig3c}
	\end{subfigure}

\hspace*{0.5cm}
	\begin{subfigure}{0.3\textwidth}
\hspace*{0.5cm}
		\includegraphics[scale=0.4]{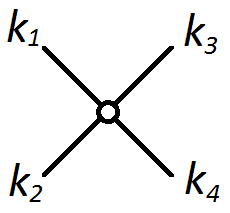}
		\caption{} \label{fig3d}
	\end{subfigure}
	\begin{subfigure}{0.3\textwidth}
\hspace*{0.5cm}
		\includegraphics[scale=0.4]{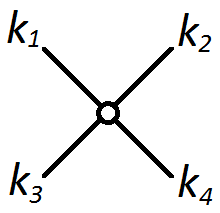}
		\caption{} \label{fig3e}
	\end{subfigure}
	\begin{subfigure}{0.3\textwidth}
\hspace*{0.5cm}
		\includegraphics[scale=0.4]{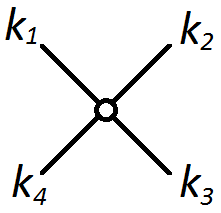}
		\caption{} \label{fig3f}
	\end{subfigure}
	\caption{\sl Four-point Green's function diagrams with one vertex} \label{fig3}
\end{figure}

\begin{figure}[htb]
\renewcommand*\thesubfigure{\arabic{subfigure}}
\makeatletter
\renewcommand{\p@subfigure}{\thefigure-}
\makeatother
\hspace*{0.5cm}
	\begin{subfigure}{0.3\textwidth}
		\includegraphics[scale=0.4]{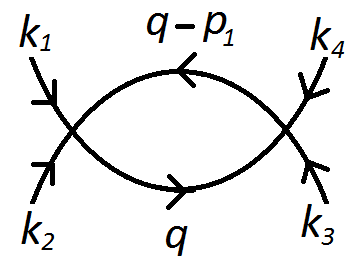}
		\caption{$p_1 = k_1+k_2$} \label{fig4a}
	\end{subfigure}
	\begin{subfigure}{0.3\textwidth}
		\includegraphics[scale=0.4]{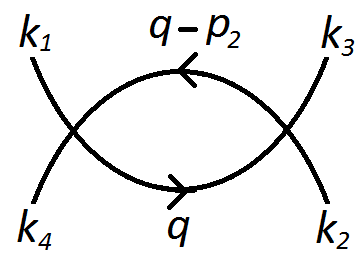}
		\caption{$p_2 = k_1+k_4$} \label{fig4b}
	\end{subfigure}
	\begin{subfigure}{0.3\textwidth}
		\includegraphics[scale=0.4]{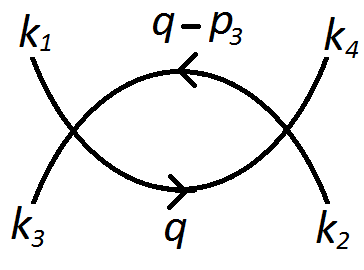}
		\caption{$p_3 = k_1+k_3$} \label{fig4c}
	\end{subfigure}

	\caption{\sl Four-point Green's function diagrams in QFT at one loop order.} \label{fig4}
\end{figure}

\begin{figure}[htb]
\renewcommand*\thesubfigure{\arabic{subfigure}}
\makeatletter
\renewcommand{\p@subfigure}{\thefigure-}
\makeatother
	\begin{subfigure}{0.24\textwidth}
		\includegraphics[scale=0.38]{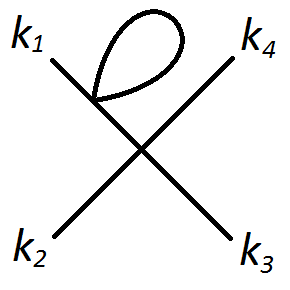}
		\caption{} \label{fig6a}
	\end{subfigure}
	\begin{subfigure}{0.24\textwidth}
		\includegraphics[scale=0.38]{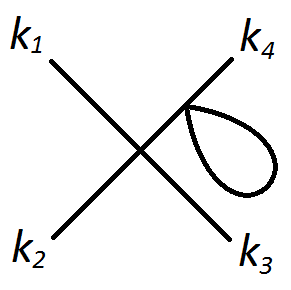}
		\caption{} \label{fig6b}
	\end{subfigure}
	\begin{subfigure}{0.24\textwidth}
		\includegraphics[scale=0.38]{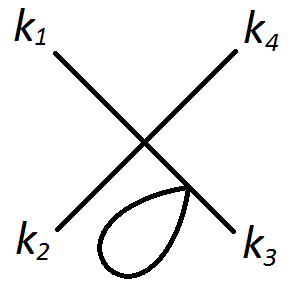}
		\caption{} \label{fig6c}
	\end{subfigure}
	\begin{subfigure}{0.24\textwidth}
		\includegraphics[scale=0.38]{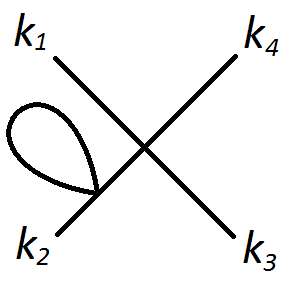}
		\caption{} \label{fig6d}
	\end{subfigure}
	\caption{\sl Four-point Green's function diagrams in QFT at one loop order} \label{fig6}
\end{figure}

\begin{figure}[htb]
\renewcommand*\thesubfigure{\arabic{subfigure}}
\makeatletter
\renewcommand{\p@subfigure}{\thefigure-}
\makeatother

	\begin{subfigure}{0.24\textwidth}
		\includegraphics[scale=0.3]{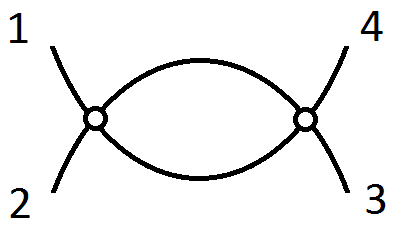}
		\caption{$\quad \begin{aligned} E_{1,1}=  k_1 \wedge k_2 \\ + k_3 \wedge k_4 \end{aligned}  $} \label{fig5.1}
	\end{subfigure}
	\begin{subfigure}{0.24\textwidth}
		\includegraphics[scale=0.3]{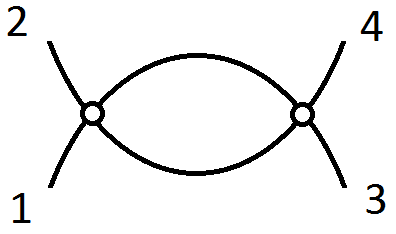}
		\caption{$\begin{aligned} F_{1,2}\,=  -k_1 \wedge k_2 \\ + k_3 \wedge k_4 \end{aligned}$} \label{fig5.2}
	\end{subfigure}
	\begin{subfigure}{0.24\textwidth}
		\includegraphics[scale=0.3]{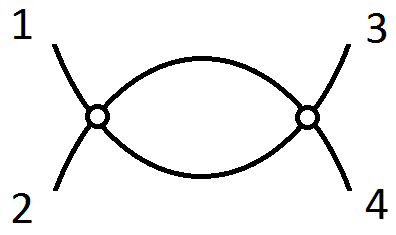}
		\caption{$\quad\begin{aligned} F_{1,3}=  k_1 \wedge k_2 \\ - k_3 \wedge k_4 \end{aligned}$} \label{fig5.3}
	\end{subfigure}
	\begin{subfigure}{0.24\textwidth}
		\includegraphics[scale=0.3]{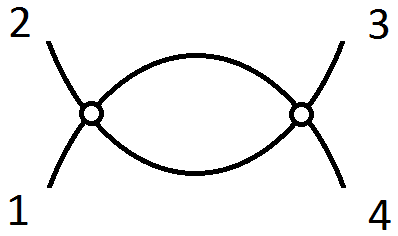}
		\caption{$\begin{aligned} F_{1,4}\,=  -k_1 \wedge k_2 \\ - k_3 \wedge k_4 \end{aligned}$} \label{fig5.4}
	\end{subfigure}

	\begin{subfigure}{0.24\textwidth}
		\includegraphics[scale=0.3]{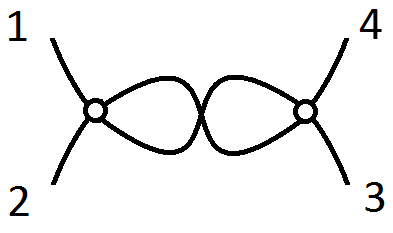}
		\caption{$\quad \begin{aligned} F_{1,5}=  k_1 \wedge k_2 \\ + k_3 \wedge k_4 \\ +2k \wedge q\end{aligned}  $} \label{fig5.5}
	\end{subfigure}
	\begin{subfigure}{0.24\textwidth}
		\includegraphics[scale=0.3]{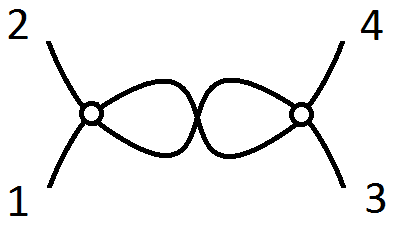}
		\caption{$ \begin{aligned} F_{1,6}\,=  -k_1 \wedge k_2 \\ + k_3 \wedge k_4 \\ +2k \wedge q\end{aligned}  $} \label{fig5.6}
	\end{subfigure}
	\begin{subfigure}{0.24\textwidth}
		\includegraphics[scale=0.3]{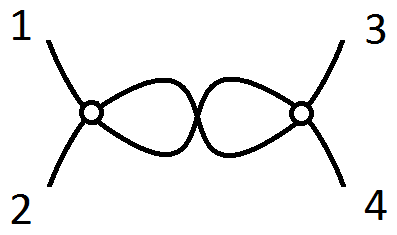}
		\caption{$\quad \begin{aligned} F_{1,7}=  k_1 \wedge k_2 \\ - k_3 \wedge k_4 \\ +2k \wedge q\end{aligned}  $} \label{fig5.7}
	\end{subfigure}
	\begin{subfigure}{0.24\textwidth}
		\includegraphics[scale=0.3]{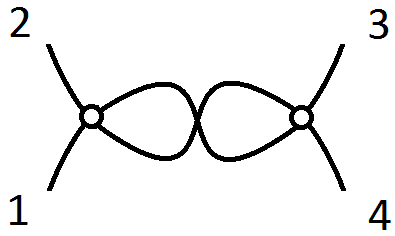}
		\caption{$ \begin{aligned} F_{1,8}\,=-  k_1 \wedge k_2 \\ - k_3 \wedge k_4 \\ +2k \wedge q\end{aligned}  $} \label{fig5.8}
	\end{subfigure}

	\begin{subfigure}{0.24\textwidth}
		\includegraphics[scale=0.3]{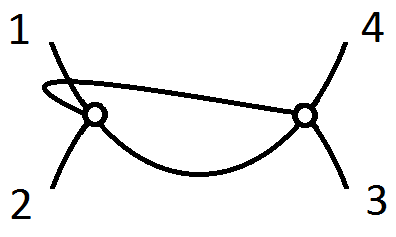}
		\caption{$ \quad \begin{aligned} F_{1,9}=  k_1 \wedge k_2 \\ + k_3 \wedge k_4 \\ -2k_1 \wedge q\end{aligned}$} \label{fig5.9}
	\end{subfigure}
	\begin{subfigure}{0.24\textwidth}
		\includegraphics[scale=0.3]{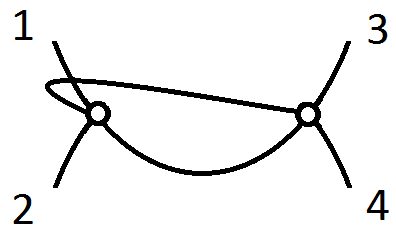}
		\caption{$ \begin{aligned} F_{1,10}=  k_1 \wedge k_2 \\ - k_3 \wedge k_4 \\ -2k_1 \wedge q\end{aligned}$} \label{fig5.10}
	\end{subfigure}
	\begin{subfigure}{0.24\textwidth}
		\includegraphics[scale=0.3]{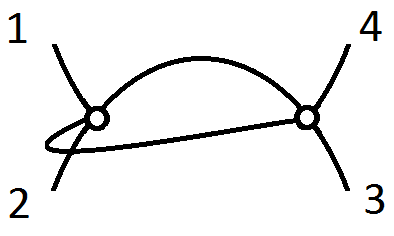}
		\caption{$ \begin{aligned} F_{1,11}=  k_1 \wedge k_2 \\ + k_3 \wedge k_4 \\ +2k_2 \wedge q\end{aligned}$} \label{fig5.11}
	\end{subfigure}
	\begin{subfigure}{0.24\textwidth}
		\includegraphics[scale=0.3]{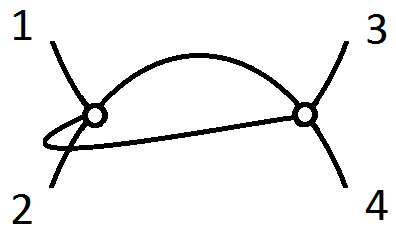}
		\caption{$ \begin{aligned} F_{1,12}=  k_1 \wedge k_2 \\ - k_3 \wedge k_4 \\ +2k_2 \wedge q\end{aligned}$} \label{fig5.12}
	\end{subfigure}

	\begin{subfigure}{0.24\textwidth}
		\includegraphics[scale=0.3]{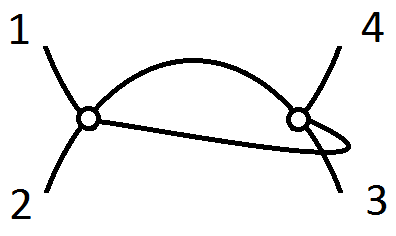}
		\caption{$ \begin{aligned} F_{1,13}= k_1 \wedge k_2 \\ + k_3 \wedge k_4 \\ +2k_3 \wedge q\end{aligned}  $} \label{fig5.13}
	\end{subfigure}
	\begin{subfigure}{0.24\textwidth}
		\includegraphics[scale=0.3]{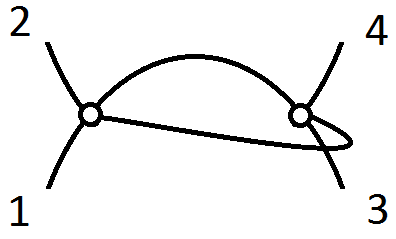}
		\caption{$ \begin{aligned} F_{1,14}=-  k_1 \wedge k_2 \\ + k_3 \wedge k_4 \\ +2k_3 \wedge q\end{aligned}  $} \label{fig5.14}
	\end{subfigure}
	\begin{subfigure}{0.24\textwidth}
		\includegraphics[scale=0.3]{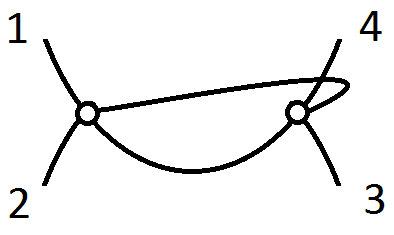}
		\caption{$ \begin{aligned} F_{1,15}=+ k_1 \wedge k_2 \\ + k_3 \wedge k_4 \\ -2k_4 \wedge q\end{aligned}  $} \label{fig5.15}
	\end{subfigure}
	\begin{subfigure}{0.24\textwidth}
		\includegraphics[scale=0.3]{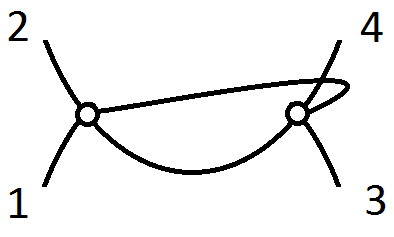}
		\caption{$ \begin{aligned} F_{1,16}=-  k_1 \wedge k_2 \\ + k_3 \wedge k_4 \\ -2k_4 \wedge q\end{aligned}  $} \label{fig5.16}
	\end{subfigure}

	\begin{subfigure}{0.5\textwidth}
\hspace*{2cm}
		\includegraphics[scale=0.3]{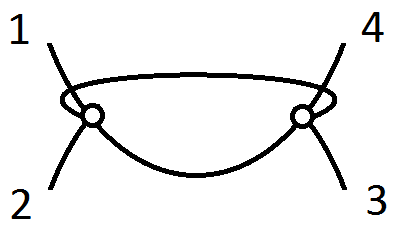}
		\caption{$ \begin{aligned} E_{1,17}=  k_1 \wedge k_2  + k_3 \wedge k_4 \\ -2(k_1+k_4) \wedge q\end{aligned}  $} \label{fig5.17}
	\end{subfigure}
	\begin{subfigure}{0.5\textwidth}
\hspace*{2cm}
		\includegraphics[scale=0.3]{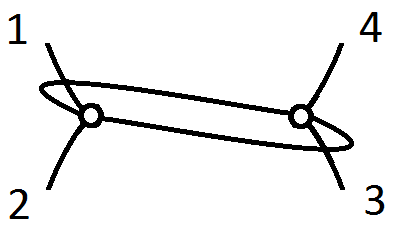}
		\caption{$ \begin{aligned} E_{1,18}=-  k_1 \wedge k_2  + k_3 \wedge k_4 \\ +2(k_1+k_3) \wedge q\end{aligned}  $} \label{fig5.18}
	\end{subfigure}

	\caption{\sl Four-point Green's function diagrams in NCQFT at one loop order with their corresponding value of $E_{nm}$. The value $n=1$ indicates that these diagrams reduce to the diagram \ref{fig4a} in the commutative case.} \label{fig5}
\end{figure}

We now compute the four-point Green's function. At first order we have as usual the diagram shown in figure \ref{fig3a},
 but in NCQFT the six diagrams shown in figure \ref{fig3} are not equivalent, so we have to consider them separately. Each of these diagrams has a symmetry factor of $4/4!$ because there are four ways to attach the first external line to the vertex, and then the other three are completely determined. Recalling that  $ \sum_{a<b}  k_a \cdot k_b=-\frac{1}{2}\sum k^2_a$ is invariant under any permutation, we can easily see that we have
 	\begin{align}
	 \tilde{G}_{\begin{NoHyper}\ref{fig3}\end{NoHyper}\text{-}j}^{(4)}(k_1,k_2,k_3,k_4)  &= -i \frac{g}{6} \frac{ e^{-\frac{1}{2} s\theta  \sum_a \boldsymbol{k}^2_a} }{\prod_a ( k_a^2 - m^2)} e^{-\frac{\theta}{2}[ -s\frac{1}{2}\sum_{a}\boldsymbol k^2_a+iE_j]} \delta^{(3)}\left( \sum_{a=1}^4 k_a \right) \nonumber \\
	&=  -i \frac{g}{6} \frac{ e^{-\frac{1}{4} s\theta  \sum_a \boldsymbol{k}^2_a-\frac{i\theta}{2} E_j}}{\prod_a ( k_a^2 - m^2)} \delta^{(3)}\left( \sum_{a=1}^4 k_a \right)  \label{48}
	\end{align}
where  $E_j$, with $j \in \{ 1,\dots 6\}$, is the Moyal part of the vertex of the corresponding diagram (see equation \eqref{34}). For example, for the diagram \ref{fig3b} we have
 	\begin{equation*}
	E_2= \boldsymbol k_1 \wedge \boldsymbol k_3 + \boldsymbol k_1 \wedge \boldsymbol k_2 + \boldsymbol k_1 \wedge \boldsymbol k_4 + \boldsymbol k_3 \wedge \boldsymbol k_2 + \boldsymbol k_3 \wedge \boldsymbol k_4 + \boldsymbol k_2 \wedge \boldsymbol k_4 .
	\end{equation*}
\renewcommand*{\thefootnote}{\fnsymbol{footnote}}At one loop order, the only connected diagrams we would have in QFT are shown in figures \ref{fig4}
 and \ref{fig6}.
 Let us first look at diagrams \ref{fig4}. In QFT, each of these diagrams has a symmetry factor of $1/2$. In NCQFT, for each of these three diagrams there are 18 different diagrams\footnote{figure \ref{fig5} shows the 18 diagrams that reduce to the diagram \ref{fig4a} in the commutative case. Equivalently, there are 18 different diagrams that reduce to the diagram \ref{fig4b} and 18 that reduce to \ref{fig4c}}, shown in figure \ref{fig5}
. The symmetry factor of each of these diagrams is equal to $(4 \cdot 4 )/(4! \cdot 4!)=1/(2\cdot 18)$. To see this, note that in this case the only symmetries of the diagrams are the cyclic permutations of the two vertices, so the first external line of each vertex has four options, and then the positions of the other three lines of each vertex are completely determined (each non-cyclic permutation gives rise to an inequivalent diagram). Note that the sum of the symmetry factors of the diagrams in figure \ref{fig5} is equal to the symmetry factor of diagram \ref{fig4a}, as expected.  

Using again the fact that  $ \sum_{a<b}  \boldsymbol k_a \cdot \boldsymbol k_b=-\frac{1}{2}\sum \boldsymbol k^2_a$ is invariant under any permutation, we can see that the non-Moyal part of the two vertices is given by 
 	\begin{equation*}
	-\frac{1}{2}( \boldsymbol{k}^2_1+ \boldsymbol{k}^2_2+ \boldsymbol{q}^2+ (\boldsymbol p_n- \boldsymbol{q})^2+ \boldsymbol{k}^2_3+ \boldsymbol{k}^2_4+ \boldsymbol{(-q)}^2+ (\boldsymbol{q}- \boldsymbol p_n)^2)
	\end{equation*}
 	\begin{equation*}
	=-\frac{1}{2} \sum_a \boldsymbol{k}^2_a- \boldsymbol q^2- (\boldsymbol p_n- \boldsymbol{q})^2 
	\end{equation*}
where the index $n\in\{1,2,3\}$ is a label for the three diagrams \ref{fig4}, and the values of $p_i$ are shown in the same figure. The last two factors of this expression cancel with the two exponentials coming from the propagators of the two internal lines. So, the corresponding integral of each of the diagrams in figure \ref{fig5} is of the form
 	\begin{align*}
	 \tilde{G}_{\begin{NoHyper}\ref{fig5}\end{NoHyper}\text{-}nm}^{(4)}(k_1&,k_2,k_3,k_4) \\= \frac{(-ig)^2}{36}& \int \frac{d^3 q}{(2\pi)^3} \frac{ e^{-\frac{1}{2} s\theta  \sum_a \boldsymbol{k}^2_a}\; e^{-\frac{\theta}{2} \left[ -s\frac{1}{2} \sum_{a}\boldsymbol k^2_a +iF_{nm} \right]}  \; \delta^{(3)}\left( \sum_{a=1}^4 k_a \right) }{(q^2-m^2)((p_n-q)^2-m^2)\prod_a ( k_a^2 - m^2)} \\
	= \frac{-g^2}{36}& \int \frac{d^3 q}{(2\pi)^3} \frac{ e^{-\frac{1}{4} s\theta \sum_a \boldsymbol k^2_a}\; e^{-i\frac{\theta}{2}F_{nm}}  \;  }{(q^2-m^2)((p_n-q)^2-m^2)\prod_a ( k_a^2 - m^2)} \delta^{(3)}\left( \sum_{a=1}^4 k_a \right)
	\end{align*}
where the first subindex $n$, as we said, is a label for the three commutative diagrams \ref{fig4}, while the second subindex $m$ is a label for the 18 noncommutative diagrams that reduce to the corresponding diagram in figure \ref{fig4}. More explicitly, $\tilde{G}_{\begin{NoHyper}\ref{fig5}\end{NoHyper}\text{-}1m}$ for $m\in \{ 1,2,\dots ,18 \}$ corresponds to the 18 diagrams (shown in figure \ref{fig5}) that reduce to \ref{fig4a} in the commutative limit, and equivalently\begin{NoHyper} $\tilde{G}_{\ref{fig5}\text{-}2m}$ and $\tilde{G}_{\ref{fig5}\text{-}3m}$\end{NoHyper}  correspond to the diagrams that reduce to \ref{fig4b} and \ref{fig4c} respectively.  $F_{nm}$ is the Moyal part of the two vertices and its value is also shown in figure \ref{fig5}.\footnote{Actually, just the values of $F_{1m}$ are shown, but the values of $F_{2m}$ and $F_{3m}$ are found in a similar way.} 

\begin{figure}[htb]
\renewcommand*\thesubfigure{\arabic{subfigure}}
\makeatletter
\renewcommand{\p@subfigure}{\thefigure-}
\makeatother
	\begin{subfigure}{0.32\textwidth}
\hspace*{0.5cm}
		\includegraphics[scale=0.35]{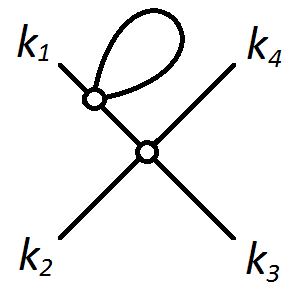}
		\caption{} \label{fig7a}
	\end{subfigure}
	\begin{subfigure}{0.32\textwidth}
\hspace*{0.5cm}
		\includegraphics[scale=0.35]{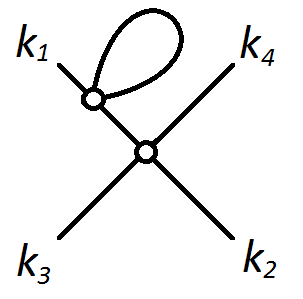}
		\caption{} \label{fig7b}
	\end{subfigure}
	\begin{subfigure}{0.32\textwidth}
\hspace*{0.5cm}
		\includegraphics[scale=0.35]{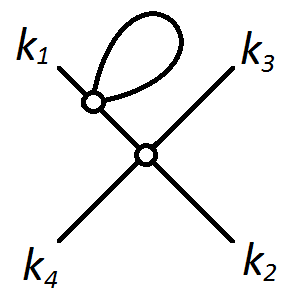}
		\caption{} \label{fig7c}
	\end{subfigure}
	\begin{subfigure}{0.32\textwidth}
\hspace*{0.5cm}
		\includegraphics[scale=0.35]{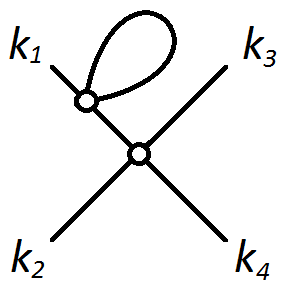}
		\caption{} \label{fig7d}
	\end{subfigure}
	\begin{subfigure}{0.32\textwidth}
\hspace*{0.5cm}
		\includegraphics[scale=0.35]{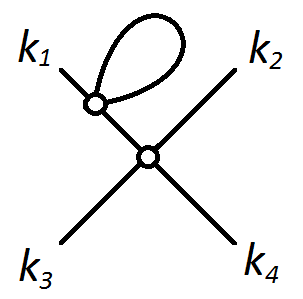}
		\caption{} \label{fig7e}
	\end{subfigure}
	\begin{subfigure}{0.32\textwidth}
\hspace*{0.5cm}
		\includegraphics[scale=0.35]{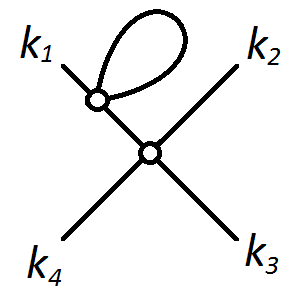}
		\caption{} \label{fig7f}
	\end{subfigure}
	\caption{\sl Four-point Green's function diagrams in NCQFT at one loop order} \label{fig7}
\end{figure}

\begin{figure}[htb]
\renewcommand*\thesubfigure{\arabic{subfigure}}
\makeatletter
\renewcommand{\p@subfigure}{\thefigure-}
\makeatother
	\begin{subfigure}{0.32\textwidth}
\hspace*{0.5cm}
		\includegraphics[scale=0.35]{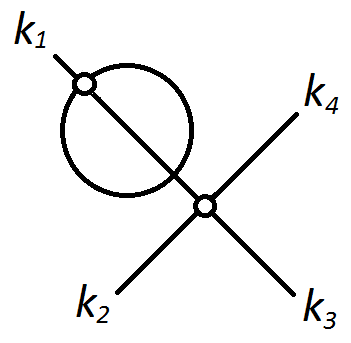}
		\caption{} \label{fig8a}
	\end{subfigure}
	\begin{subfigure}{0.32\textwidth}
\hspace*{0.5cm}
		\includegraphics[scale=0.35]{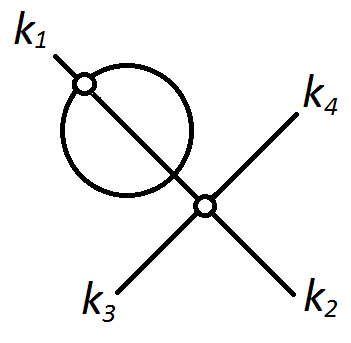}
		\caption{} \label{fig8b}
	\end{subfigure}
	\begin{subfigure}{0.32\textwidth}
\hspace*{0.5cm}
		\includegraphics[scale=0.35]{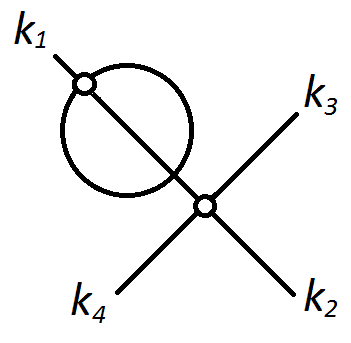}
		\caption{} \label{fig8c}
	\end{subfigure}
	\begin{subfigure}{0.32\textwidth}
\hspace*{0.5cm}
		\includegraphics[scale=0.35]{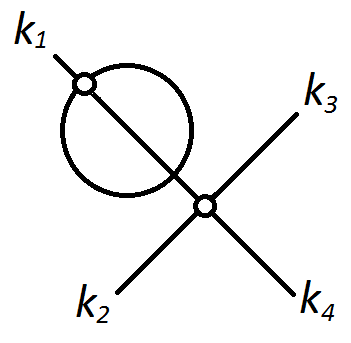}
		\caption{} \label{fig8d}
	\end{subfigure}
	\begin{subfigure}{0.32\textwidth}
\hspace*{0.5cm}
		\includegraphics[scale=0.35]{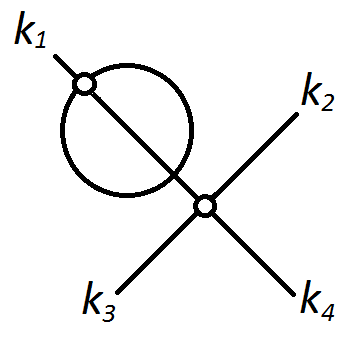}
		\caption{} \label{fig8e}
	\end{subfigure}
	\begin{subfigure}{0.32\textwidth}
\hspace*{0.5cm}
		\includegraphics[scale=0.35]{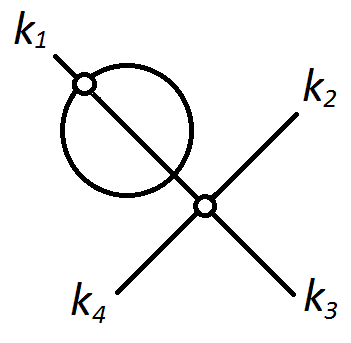}
		\caption{} \label{fig8f}
	\end{subfigure}
	\caption{\sl Four-point Green's function diagrams in NCQFT at one loop order} \label{fig8}
\end{figure}

Finally, let us look at diagrams \ref{fig6}. In NCQFT, for each of the diagrams shown in figure \ref{fig6} there are 12 different diagrams\footnote{Figures \ref{fig7} and \ref{fig8} show the 12 diagrams that reduce to diagram \ref{fig6a} in the commutative limit. The other diagrams are the same except that the loop is attached to another external line.}, shown in figures \ref{fig7}
 and \ref{fig8}.
 The diagrams in figure \ref{fig7} have a symmetry factor of $(4\cdot4\cdot2)/(4!\cdot4!)=1/18$ and their corresponding integrals are given by  
 	\begin{align*}
	 \tilde{G}_{\begin{NoHyper}\ref{fig7}\end{NoHyper}\text{-}ij}^{(4)}&(k_1,k_2,k_3,k_4) \\
	&= \frac{(-ig)^2}{18} \int \frac{d^3 q}{(2\pi)^3} \frac{ e^{-\frac{1}{4} s\theta \sum_a \boldsymbol{k}^2_a }\; e^{-i\frac{\theta}{2} E_j }  \;  }{(q^2-m^2)(k_i^2-m^2)\prod_a ( k_{a}^2 - m^2)} \delta^{(3)}\left( \sum_{a=1}^4 k_a \right)
	\end{align*}
where the $E_j$ is the same we had in equation (\ref{48}), and again the subindex $i\in\{1,\dots,4 \}$  is a label for the four commutative diagrams in figure \ref{fig6}, while the index $j\in\{1,\dots ,6 \}$ is a label for the six noncommutative diagrams (of the form shown in figure \ref{fig7}) that reduce to the corresponding commutative diagram in figure  \ref{fig6} (i.e. $ \tilde{G}_{\begin{NoHyper}\ref{fig7}\end{NoHyper}\text{-}ij}$ for  $j\in\{1,\dots ,6 \}$ correspond to the 6 noncommutative diagrams that reduce to \ref{fig6}-i in the commutative limit)

Finally, the diagrams in figure \begin{NoHyper}\ref{fig8}\end{NoHyper} have a symmetry factor of $(4\cdot4)/(4!\cdot4!)=1/36$ and their corresponding integrals are given by  
 	\begin{align*}
	 \tilde{G}_{\begin{NoHyper}\ref{fig8}\end{NoHyper}\text{-}ij}^{(4)}&(k_1,k_2,k_3,k_4) \\
	&= \frac{(-ig)^2}{36} \int \frac{d^3 q}{(2\pi)^3} \frac{e^{-i \theta \boldsymbol k_i \wedge \boldsymbol q} e^{-\frac{1}{4} s\theta  \sum_a \boldsymbol{k}^2_a }\; e^{-i\frac{\theta}{2} E_j }  \;  }{(q^2-m^2)(k_i^2-m^2)\prod_a ( k_{a}^2 - m^2)} \delta^{(3)}\left( \sum_{a=1}^4 k_a \right)
	\end{align*}
where  $E_j$ is the same as before, as well as the explanation of the subindices. We can finally write the connected four-point Green's function up to one loop order 
 	\begin{align*}
	 \tilde{G}_{c;2}^{(4)}(k_1,...,&k_4)   =\sum_{j=1}^6 \tilde{G}_{\begin{NoHyper}\ref{fig3}\end{NoHyper}\text{-}j}^{(4)} +  \sum_{n=1}^3\sum_{m=1}^{18}\tilde{G}_{\begin{NoHyper}\ref{fig5}\end{NoHyper}\text{-}nm}^{(4)} +\sum_{i=1}^4\sum_{j=1}^6\tilde{G}_{\begin{NoHyper}\ref{fig7}\end{NoHyper}\text{-}ij}^{(4)} +  \sum_{i=1}^4\sum_{j=1}^6\tilde{G}_{\begin{NoHyper}\ref{fig8}\end{NoHyper}\text{-}ij}^{(4)}\\ 
	=-\Bigg(& \sum_j \frac{ig}{6} \frac{ e^{-\frac{1}{4} s\theta  \sum_a \boldsymbol k^2_a-\frac{i\theta}{2} E_j}}{\prod_{a=1}^4 ( k_a^2 - m^2)} \\
	+& \sum_{nm}\frac{g^2}{36} \int \frac{d^3 q}{(2\pi)^3} \frac{ e^{-\frac{1}{4} s\theta \sum_a \boldsymbol k^2_a-i\frac{\theta}{2}F_{nm}}  \;  }{(q^2-m^2)((p_n-q)^2-m^2)\prod_a ( k_a^2 - m^2)}  \\
	+& \sum_{ij}\frac{g^2}{36} \int \frac{d^3 q}{(2\pi)^3} \frac{(2+e^{-i \theta \boldsymbol k_i \wedge \boldsymbol q}) e^{-\frac{1}{4} s\theta \sum_a \boldsymbol k^2_a-i\frac{\theta}{2} E_j }  \;  }{(q^2-m^2)(k_i^2-m^2)\prod_a ( k_{a}^2 - m^2)} \Bigg) \delta^{(3)}\left( \sum k_a \right).
	\end{align*}
It is not difficult to see that if we set $\theta=0$ we get back the usual expression for the connected four-point Green's function in QFT.

\section{Ultraviolet/Infrared mixing}

The Ultraviolet/Infrared mixing is a phenomenon that appears in NCQFT but it is not present in ordinary Quantum Field Theory. To understand it, let us look at equations \eqref{50} and \eqref{51}. The integral in equation \eqref{50} corresponds to diagram \ref{fig1a}, and it diverges for big $q$, like in QFT, so it has ultraviolet divergence. On the other hand, the integral in equation \eqref{51} corresponds to diagram \ref{fig1b}, but note that the integrand has an oscillating factor that softens the ultraviolet divergence but at the same time it is responsible for an infrared divergence. The same phenomenon is observed for the diagrams of the four-point Green's function. This is what is usually called ultraviolet/infrared mixing. 
Note that it is the Moyal part of the vertex what produces this mixing, that's why it is a common characteristic of the s-ordered product for any value of the parameter $s$. In the next chapter we will see that this is actually common for all the translation invariant star products.

\chapter{Field Theory with Translation Invariant Star Products} \label{chap-transinv}

In the first two chapters we studied the s-ordered star product which are a special case of the translation invariant star product. In this chapter we will introduce the general translation invariant star products and we will do some field theory with these products. 

\section{The general translation invariant star products (TISP)}

In this section all vectors are in the two dimensional space\footnote{All the expressions in this section are actually valid in any dimension} as in the first chapter, so we will omit the boldface notation, we will come back to space-time in the next section. Consider the generalization of the s-ordered product \eqref{100} given by 
 	\begin{equation*}
	(f \star g)(x)= \int \frac{d^2p}{(2\pi)^2} \frac{d^2q}{(2\pi)^2} \frac{d^2r}{(2\pi)^2} \tilde f(q) \tilde g(r) K(p,q,r) e^{-ip \cdot x}
	\end{equation*}
where $K$ is a distribution. Note that this includes the commutative products. Indeed, if we impose the condition   
 	\begin{align*}
	(f \star g)(x)&= \int \frac{d^2p}{(2\pi)^2} \frac{d^2q}{(2\pi)^2} \frac{d^2r}{(2\pi)^2} \tilde f(q) \tilde g(r) K(p,q,r) e^{-ip \cdot x} \\
	&= \int \frac{d^2p}{(2\pi)^2} \frac{d^2q}{(2\pi)^2} \frac{d^2r}{(2\pi)^2} \tilde g(q) \tilde f(r) K(p,q,r) e^{-ip \cdot x}=(g \star f)(x)
	\end{align*}
we get the commutativity condition $K(p,q,r) = K(p,r,q)$. In particular, the ordinary product is of this form with $K(p,q,r) =(2\pi)^2 \delta^{(3)}(r-p+q)$. 

Now, we want a translation invariant product, so we have to impose the condition  
 	\begin{equation*}
	\mathcal T_a (f \star g)= \mathcal T_a(f) \star \mathcal T_a(g)
	\end{equation*}
where $\mathcal T_a (f)(x) = f(x+a)$ is the translation by any vector $a \in \mathbb R^3$. The left hand side is given by 
 	\begin{equation*}
	\mathcal T_a (f \star g)= \int \frac{d^2p}{(2\pi)^2} \frac{d^2q}{(2\pi)^2} \frac{d^2r}{(2\pi)^2} \tilde f(q) \tilde g(r) K(p,q,r) e^{-ip \cdot (x+a)}
	\end{equation*}
while the right hand side is 
 	\begin{align*}
	\mathcal T_a(f) \star \mathcal T_a(g) = \int \frac{d^2p}{(2\pi)^2} \frac{d^2q}{(2\pi)^2} \frac{d^2r}{(2\pi)^2} \widetilde { \mathcal T_a (f)}(q) \widetilde { \mathcal T_a (g)}(r) K(p,q,r) e^{-ip \cdot x} \\
	 = \int \frac{d^2p}{(2\pi)^2} \frac{d^2q}{(2\pi)^2} \frac{d^2r}{(2\pi)^2} e^{ia \cdot q} \tilde f(q) e^{ia \cdot r} \tilde g(r) K(p,q,r) e^{-ip \cdot x}.
	\end{align*}
So the product is translation invariant if these two expressions are equal for any vector $a$, which means
 	\begin{equation*}
	  K(p,q,r) =(2\pi)^2 e^{\alpha (p,q)} \delta^{(2)}(r-p+q)
	\end{equation*}
where $\alpha$ is a generic (possibly complex) function. So the general translation invariant star product is given by
 	\begin{equation}
	(f \star g)(x)= \int \frac{d^2p}{(2\pi)^2} \frac{d^2q}{(2\pi)^2} \tilde f(q) \tilde g(p-q) e^{\alpha (p,q)} e^{-ip \cdot x}. \label{36}
	\end{equation}
In particular, it is not difficult to see that the ordinary product is given by $\alpha=0$ and the s-ordered product is given by 
 	\begin{equation*}
	  \alpha_s (p,q) = \frac{i \theta}{2} p\wedge q - \frac{s\theta}{2} (p-q)\cdot q 
	\end{equation*}
and clearly the Moyal product ($s=0$) is given by $ \alpha_M (p,q) = \frac{i \theta}{2} p\wedge q$, so we will call this the Moyal part of $\alpha$.

There are three constraints that the function $\alpha$ has to satisfy. The first one is the associativity condition
 	\begin{equation*}
	 ((f \star g) \star h)(x) =  (f \star (g \star h))(x) .
	\end{equation*}
A straightforward computation gives 
 	\begin{align*}
	 ((f \star g) \star h)(x) =   \int \frac{d^2p}{(2\pi)^2} \frac{d^2q}{(2\pi)^2} \frac{d^2t}{(2\pi)^2} \tilde f(t) \tilde g(q-t) \tilde h(p-q) e^{\alpha (q,t)+\alpha (p,q)} e^{-ip \cdot x}
	\end{align*}
and on the other hand
 	\begin{align*}
	 (f \star (g \star h))(x) = \int \frac{d^2p}{(2\pi)^2} \frac{d^2q}{(2\pi)^2}  \frac{d^2t}{(2\pi)^2} \tilde f(t) \tilde g(q-t)  h(p-q)  e^{\alpha (p-t,q-t)+\alpha (p,t)} e^{-ip \cdot x}.
	\end{align*}
So the associativity condition says
 	\begin{equation}
	\alpha (q,t)+\alpha (p,q)=\alpha (p-t,q-t)+\alpha (p,t). \label{70}
	\end{equation}

The second constraint is the condition that the constant function $1$ be the identity of the algebra of functions with  product  \eqref{36}, i.e.
 	\begin{equation}
	(f \star 1)(x) =\int \frac{d^2p}{(2\pi)^2} \tilde f(q) e^{\alpha (p,p)} e^{-ip \cdot x}=f \label{73}
	\end{equation}
and

 	\begin{equation}
	(1 \star f)(x) =\int \frac{d^2p}{(2\pi)^2} \tilde f(q) e^{\alpha (p,0)} e^{-ip \cdot x}=f \label{74}
	\end{equation}
which means
 	\begin{equation*}
	\alpha (p,p) =0 \quad \text{and} \quad \alpha (p,0)=0.
	\end{equation*}

Finally, the third constraint is the condition that the algebra of functions with the star product be a $*$-algebra. That is, there must be a map which satisfies the following conditions:
 	\begin{align*}
	(f^*)^* &= f \\
	(\lambda f+ \mu g)^* &= \bar \lambda f^* + \bar \mu g^* \\
	( f \star g)^* &=   g^*  \star f^*
	\end{align*}
where $\lambda,\mu \in \mathbb C$, and the bar denotes complex conjugation. In this case the involution $*$ is given by complex conjugation and just the last condition imposes a constraint on $\alpha$. The left hand side of the equation is
 	\begin{equation*}
	( f \star g)^* = \int \frac{d^2p}{(2\pi)^2} \frac{d^2q}{(2\pi)^2} \tilde f(q)^* \tilde g(p-q)^* e^{\alpha (p,q)^*} e^{ip \cdot x} 
	\end{equation*}
and the right hand side is
 	\begin{eqnarray*}
	 g^*  \star f^* &=& \int \frac{d^2p}{(2\pi)^2} \frac{d^2q}{(2\pi)^2} \widetilde {g^*}(q) \widetilde {f^*}(p-q) e^{\alpha (p,q)} e^{-ip \cdot x} \\
	 &=& \int \frac{d^2p}{(2\pi)^2} \frac{d^2q}{(2\pi)^2} \tilde g(-q)^* \tilde f(q-p)^* e^{\alpha (p,q)} e^{-ip \cdot x} \\
	 &=& \int \frac{d^2p}{(2\pi)^2} \frac{d^2q}{(2\pi)^2} \tilde f(q)^* \tilde g(p-q)^* e^{\alpha (-p,q-p)} e^{ip \cdot x},
 	\end{eqnarray*}
so $\alpha$ has to satisfy 
 	\begin{equation*}
	 \alpha(p,q)^*=\alpha(-p,q-p).
	\end{equation*}
This is a reasonable condition if we want to represent the functions as operators once we study the field theory of noncommutative spaces corresponding to these star products.  

So we have the following constraints on $\alpha$
 	\begin{eqnarray*}
		\alpha (q,t)+\alpha (p,q)&=&\alpha (p-t,q-t)+\alpha (p,t) \\
		\alpha (p,p) &=&0 \\
		\alpha (p,0)&=&0 \\
		 \alpha(p,q)^*&=&\alpha(-p,q-p)
 	\end{eqnarray*}
It has been shown that the most general function that satisfies these conditions is of the form \cite{ArdalanSadoogi}
 	\begin{equation}
	 \alpha(p,q) = \eta(q) - \eta(p) + \eta(p-q) +  i\omega(p,q) \label{37}
	\end{equation}
where $\omega(p,q) = \frac{\theta}{2} p\wedge q$ with $\theta$ an arbitrary real constant, and  $\eta(p)=\eta_1 (p) + i\eta_2 (p)$ with $\eta_1$ an arbitrary real and even function  such that $\eta_1(0)=0$, and $\eta_2$ a real odd function without linear term, i.e.
 	\begin{equation*}
	 \eta_2(p) = \sum_{n=1}^{\infty} \sum_{l=0}^{2n+1} C_{l,2n+1-l} \, p_1^l p_2^{2n+1-l}.
	\end{equation*}
Note that the $\eta$ function corresponding to the s-product is given by $\eta_2(p) = 0$ and $\eta_1 (p)=\frac{s \theta}{4} p^2$ .

Let us look at the commutativity condition $f \star g = g \star f$. The product is commutative if
 	\begin{equation*}
	(f \star g)(x) = \int \frac{d^2p}{(2\pi)^2} \frac{d^2q}{(2\pi)^2} \tilde f(q) \tilde g(p-q) e^{\alpha (p,q)} e^{-ip \cdot x}  
	\end{equation*}
equals
  	\begin{eqnarray*}
		(g \star f)(x) &=& \int \frac{d^2p}{(2\pi)^2} \frac{d^2q}{(2\pi)^2} \tilde g(q) \tilde f(p-q) e^{\alpha (p,q)} e^{-ip \cdot x} \\
		&=&\int \frac{d^2p}{(2\pi)^2} \frac{d^2q}{(2\pi)^2} \tilde f(q) \tilde g(p-q) e^{\alpha (p,p-q)} e^{-ip \cdot x} 
 	\end{eqnarray*}
which means
 	\begin{equation*}
	\alpha(p,q) = \alpha (p,p-q)
	\end{equation*}
but note that, from \ref{37} we have
 	\begin{equation*}
	\alpha(p,q) =  \eta(q) - \eta(p) + \eta(p-q) +  i\omega(p,q)
	\end{equation*}
and
 	\begin{equation*}
	\alpha(p,p-q) =  \eta(q) - \eta(p) + \eta(p-q) +  i\omega(p,p-q).
	\end{equation*}
Therefore, just $\omega$ contributes to the non-commutativity of the product. But note also that $i\omega(p,q)=\alpha_M(p,q) $, so it is just the Moyal part of $\alpha$ what contributes to the non-commutativity of the product. Heuristically, any noncommutative TISP is the moyal product ``plus" some (non-local) commutative product, while any commutative TISP is of the form $\alpha(p,p-q) =  \eta(q) - \eta(p) + \eta(p-q)$. This also implies that any noncommutative TISP satisfies the commutator $[{x}^i,{x}^j]_{\star} = i \theta^{ij}$.   

\section{Quantum Field Theory with a general translation invariant star product}
Let us now discuss the noncommutative field theory with a general translation invariant star product. Note that we are back to the $(2+1)$-dimensional space-time. We consider again the action \eqref{7}
	\begin{equation}
	S = \int d^3x \left( \frac{1}{2}\partial_\mu \phi \star \partial^\mu \phi - \frac{m^2}{2} \phi \star \phi - \frac{g}{4!} \phi \star \phi \star \phi \star \phi \right).
	\end{equation}
To compute the equation of motion we first need the following property
 	\begin{eqnarray}
	\int d^3 x \, f \star g &=& \int d^3 x \, \frac{d^3p}{(2\pi)^3}\frac{d^3q}{(2\pi)^3} \tilde f(q) \tilde g(p-q) e^{\alpha (p,q)} e^{-ip \cdot x} \nonumber \\
	&=&  \int \frac{d^3q}{(2\pi)^3} \tilde f(q) \tilde g(-q) e^{\alpha (0,q)}, \label{40}
	\end{eqnarray}
and then we proceed as in section \ref{classfieldt}. Using this property and equation \eqref{38} (which is also valid for the general translation invariant product), we have
 	\begin{equation*}
	\delta S_0 = - \int \frac{d^3q}{(2\pi)^3} \widetilde {\delta \phi}(q)  (-q^2 + m^2)\tilde \phi(-q) e^{\alpha (0,q)} 
	\end{equation*}
for any variation of the field $\delta \phi$. So the equation of motion is 
 	\begin{equation*}
	 e^{\alpha (0,-q)}  (q^2 - m^2)\tilde \phi(q) = 0 
	\end{equation*}
or equivalently
 	\begin{equation}
	 e^{2\eta_1(q)}  (q^2 - m^2)\tilde \phi(q) = 0 \label{39}
	\end{equation}
which again reduces to the same ordinary equation of motion due to the invertibility of the exponential factor. So, as for the s-ordered product, we find that at the classical level, the noncommutative free field theory given by the action \eqref{7}, is the same as the commutative one.

We now proceed to the computation of the propagator and the vertex. Te propagator can be easily found from equation \eqref{39} to be 
 	\begin{equation}
	 \tilde G_0 (p) = \frac{ e^{-2\eta_1(q)} }{ (p^2 - m^2)}.
	\end{equation}
The vertex can be computed using equations \eqref{36} and \eqref{40}
 	\begin{align*}
	 S_{\text{int}} =& \frac{g}{4!} \int d^3x \frac{d^3k_1}{(2\pi)^3}\frac{d^3k_2}{(2\pi)^3}\frac{d^3k_3}{(2\pi)^3}\frac{d^3k_4}{(2\pi)^3} \tilde \phi(k_2) \tilde \phi(k_1-k_2) \tilde 	 \phi(k_4) \tilde \phi(k_3-k_4) \nonumber \\ &  \hspace{16em}   e^{\alpha(k_1,k_2)}e^{\alpha(k_3,k_4)}e^{k_1 \cdot x}\star e^{k_3 \cdot x} \nonumber \\
	  = &\frac{g}{4!} \int \frac{d^3k_1}{(2\pi)^3}\frac{d^3k_2}{(2\pi)^3}\frac{d^3k_3}{(2\pi)^3}\frac{d^3k_4}{(2\pi)^3} \tilde \phi(k_2) \tilde \phi(k_1-k_2) \tilde \phi(k_4) \tilde \phi(k_3-k_4) \nonumber \\ & \qquad e^{\alpha(k_1,k_2)}e^{\alpha(k_3,k_4)}  \int \frac{d^3k}{(2\pi)^3}e^{\alpha(0,k)}(2\pi)^3 \delta^{(3)}(k_1 - k) (2\pi)^3 \delta^{(3)}(k_3 + k)  \nonumber \\
	  = &\frac{g}{4!}(2\pi)^3 \int \frac{d^3k_1}{(2\pi)^3}\frac{d^3k_2}{(2\pi)^3}\frac{d^3k_3}{(2\pi)^3}\frac{d^3k_4}{(2\pi)^3} \tilde \phi(k_2) \tilde \phi(k_1-k_2) \tilde \phi(k_4) \tilde \phi(k_3-k_4) \nonumber \\ & \hspace{12em}  e^{\alpha(k_1,k_2) + \alpha(k_3,k_4) + \alpha(0,k_1)} \delta^{(3)}(k_1 + k_3) 
	\end{align*}
which can be written as 
 	\begin{eqnarray*}
	 S_{\text{int}}  = \frac{g}{4!}(2\pi)^3 \int \frac{d^3k_1}{(2\pi)^3}\frac{d^3k_2}{(2\pi)^3}\frac{d^3k_3}{(2\pi)^3}\frac{d^3k_4}{(2\pi)^3} \tilde \phi(k_1) \tilde \phi(k_2) \tilde \phi(k_3) \tilde \phi(k_4) \nonumber \\ e^{\alpha(k_1+k_2,k_2) + \alpha(k_3+k_4,k_4) + \alpha(0,k_1+k_2)} \delta^{(3)}(k_1 +k_2+ k_3+k_4) 
	\end{eqnarray*}
So the vertex is given by
 	\begin{equation}
	V_{\star} = V e^{\alpha(k_1+k_2,k_2) + \alpha(k_3+k_4,k_4) + \alpha(0,k_1+k_2)} \label{41}
	\end{equation}
where $V$ is again the ordinary vertex given in equation \eqref{101}. Note that the vertex \eqref{41} can be written, using equation \eqref{37}, as
 	\begin{align*}
	V_{\star} = V e^{\eta(k_1) - \eta(k_1 + k_2) + \eta(k_2) +  i\omega(k_1+k_2,k_1) } e^{ \eta(k_3) - \eta(k_3 + k_4) + \eta(k_4) +  i\omega(k_3+k_4,k_3) } \\  e^{ \eta(k_1+k_2) - \eta(0) + \eta(-k_1-k_2) +  i\omega(0,k_1+k_2) }
	\end{align*}
 	\begin{equation*}
	= V e^{\eta(k_1) + \eta(k_2)+ \eta(k_3) + \eta(k_4) +  i\omega(k_2,k_1)  - \eta(k_3 + k_4)  +  i\omega(k_4,k_3) + \eta(-k_1-k_2) }
	\end{equation*}
But using conservation of momentum in the vertex $-k_1-k_2 = k_3 + k_4$ and recalling that $ \omega(p,q) =  \frac{\theta}{2} p\wedge q $ we have $- \eta(k_3 + k_4)+ \eta(-k_1-k_2)=0$ and $ \omega(k_2,k_1)  +\omega(k_4,k_3) =- \sum_{a<b} \omega(k_a,k_b)$, so the vertex can be written as
 	\begin{equation}
	V_{\star}= V e^{\sum_a \eta(k_a) -   i\frac{\theta}{2} \sum_{a<b} k_a \wedge k_b} \label{42}
	\end{equation}
%
\section{Green's functions for a general translation invariant star product}

It is important to note that the factor $\sum_a \eta(k_a)$ in the vertex \eqref{42} is invariant under any permutation, so the symmetries of the vertices depend just on the Moyal part $ i\frac{\theta}{2} \sum_{a<b} k_a \wedge k_b$. This means that if we want to do Noncommutative Quantum Field Theory with a translation invariant star product with non-zero Moyal part, then we know that all the diagrams and symmetry factors at any order, are the same for any of these products. So the diagrams of the two-point and four-point Green's functions are the ones we found in sections \ref{2pointgf} and \ref{4pointgf}. 

Note that it is possible to do Quantum Field Theory with a translation invariant product without Moyal part. But we saw that just the Moyal part contributes to the non-commutativity of the product. This means that a translation invariant product without Moyal part would give rise to a commutative (possibly non-local) Quantum Field Theory, in which case we would have the same diagrams as in conventional QFT. But here we are interested in the noncommutative products, which give rise to the commutator \eqref{3}. So we will only consider the star products with non-zero Moyal part (i.e noncommutative) and we will call them NCTISP.

We now proceed to the computation of the two-point and four-point  Green's functions. Let us begin by the two-point Green's function $\tilde G^{(2)}(p) $. At leading order it is, as usual,  given by the propagator 
 	\begin{equation}
	\tilde G_0^{(2)}(p) = \tilde G_0 (p) = \frac{e^{-2\eta_1(q)} }{ (p^2 - m^2)}.
	\end{equation}
At first order, there are two diagrams \ref{fig1a} and \ref{fig1b}. The Moyal part is the same we found in sections \ref{2pointgf} and \ref{4pointgf}. The non-Moyal part of the vertex is given by   
 	\begin{equation}
	e^{\eta(p)+\eta(-p)+\eta(q)+\eta(-q) } = e^{2(\eta_1(p)+\eta_1(q)) }.
	\end{equation}
The integral corresponding to diagram \ref{fig1a} is then given by 
	\begin{align*}
	 \tilde{G}_{\begin{NoHyper}\ref{fig1a}\end{NoHyper}}^{(2)}(p)&=\frac{-ig}{3} \int \frac{d^3 q}{(2\pi)^3}\, \frac{ e^{-( 4\eta_1(p)+ 2\eta_1(q))}  e^{2(\eta_1(p)+\eta_1(q)) } }{ (p^2 - m^2)^2 (q^2 - m^2)} \\
	&= \frac{-ig}{3} \int \frac{d^3 q}{(2\pi)^3}\, \frac{  e^{- 2\eta_1(p)} }{ (p^2 - m^2)^2 (q^2 - m^2)} 
	\end{align*}
and the one corresponding to diagram \ref{fig1b} is
	\begin{equation*}
	  \tilde{G}_{\begin{NoHyper}\ref{fig1b}\end{NoHyper}}^{(2)}(p)= \frac{-ig}{6} \int \frac{d^3 q}{(2\pi)^3}\, \frac{  e^{- 2\eta_1(p) - i\theta \boldsymbol p\wedge \boldsymbol q} }{ (p^2 - m^2)^2 (q^2 - m^2)} 
	\end{equation*}
The connected two-point Green's function at one loop order is finally given by 
 	\begin{equation*}
	\tilde G_{c; \,1}^{(2)}(p) =  \frac{e^{-2\eta_1(q)} }{ (p^2 - m^2)}- \frac{ig}{6} \int \frac{d^3 q}{(2\pi)^3}\, \frac{( 2\, +  e^{- \theta i \boldsymbol p\wedge \boldsymbol q}) e^{- 2\eta_1(p)}  }{  (p^2 - m^2)^2 (q^2 - m^2)} .
	\end{equation*}
Note that if we substitute the $\eta$ function corresponding to the s-ordered function ($\eta_2(p) = 0$ and $\eta_1 (p)=\frac{s \theta}{4} p^2$), we get back the expression \eqref{45}.

Let us compute the four-point Green's function. At first order, the non-Moyal part of the vertex is given by $\text{exp}(\sum_a \eta(k_a))$, so we have  
 	\begin{align}
	 \tilde{G}_{\begin{NoHyper}\ref{fig3}\end{NoHyper}\text{-}j}^{(4)}(k_1,k_2,k_3,k_4)  &= -i \frac{g}{6} \frac{ e^{-2\sum_a \eta_1 (k_a)} }{\prod_a ( k_a^2 - m^2)} e^{\sum_a \eta(k_a)-i\frac{\theta}{2}E_j} \delta^{(3)}\left( \sum_{a=1}^4 k_a \right) \nonumber \\
	&=  -i \frac{g}{6} \frac{ e^{-\sum_a \bar \eta(k_a)-\frac{i\theta}{2} E_j}}{\prod_a ( k_a^2 - m^2)} \delta^{(3)}\left( \sum_{a=1}^4 k_a \right)  
	\end{align}
where $E_j$ is the same we had in section \ref{4pointgf}, and $\bar \eta$ is the complex conjugate of $\eta$. So, now we move to second order. The non-Moyal part of the diagrams of the form \ref{fig5} is 
 	\begin{equation*}
	 \eta (k_1)+ \eta (k_2)+ \eta (q)+ \eta(p_n- q)+ \eta (k_3)+ \eta (k_4)+ \eta (-q)+ \eta (q- p_n)
	\end{equation*}
 	\begin{equation*}
	= \sum_a \eta (k_a)+ 2\eta_1 (q)+ 2\eta_1(p_n- q) 
	\end{equation*}
and again, as with the s-ordered product, the last two factors of this expression cancel with the exponentials of the two propagators of the internal lines. So for diagrams of the form \ref{fig5} we have
 	\begin{align*}
	 \tilde{G}&_{\begin{NoHyper}\ref{fig5}\end{NoHyper}\text{-}nm}^{(4)}(k_1,k_2,k_3,k_4) \\&= \frac{(-ig)^2}{36} \int \frac{d^3 q}{(2\pi)^3} \frac{ e^{-2\sum_a \eta_1 (k_a)}\; e^{\sum_a \eta (k_a) -i\frac{\theta}{2} F_{nm}}  \; \delta^{(3)}\left( \sum_{a=1}^4 k_a \right) }{(q^2-m^2)((p_n-q)^2-m^2)\prod_a ( k_a^2 - m^2)} \\
	&=\frac{-g^2}{36} \int \frac{d^3 q}{(2\pi)^3} \frac{ e^{-\sum_a \bar\eta(k_a)-i\frac{\theta}{2} F_{nm}} }{(q^2-m^2)((p_n-q)^2-m^2)\prod_a ( k_a^2 - m^2)}\delta^{(3)}\left( \sum_{a=1}^4 k_a \right)
	\end{align*}
where $F_{nm}$ is the same we had in section \ref{4pointgf}. In a similar way, it is easy to see that for the diagrams of the form \ref{fig7} and \ref{fig8} we have

 	\begin{align*}
	 \tilde{G}_{\begin{NoHyper}\ref{fig7}\end{NoHyper}\text{-}ij}^{(4)}&(k_1,k_2,k_3,k_4) \\
	&= \frac{(-ig)^2}{18} \int \frac{d^3 q}{(2\pi)^3} \frac{ e^{-\sum_a \bar \eta(k_a)-i\frac{\theta}{2} E_j }  \;  }{(q^2-m^2)(k_i^2-m^2)\prod_a ( k_{a}^2 - m^2)} \delta^{(3)}\left( \sum_{a=1}^4 k_a \right)
	\end{align*} 
and
 	\begin{align*}
	 \tilde{G}_{\begin{NoHyper}\ref{fig8}\end{NoHyper}\text{-}ij}^{(4)}&(k_1,k_2,k_3,k_4) \\
	&= \frac{(-ig)^2}{36} \int \frac{d^3 q}{(2\pi)^3} \frac{ e^{-i \theta \boldsymbol k_i \wedge \boldsymbol q}\, e^{-\sum_a \bar \eta(k_a)-i\frac{\theta}{2} E_j }  \;  }{(q^2-m^2)(k_i^2-m^2)\prod_a ( k_{a}^2 - m^2)} \delta^{(3)}\left( \sum_{a=1}^4 k_a \right)
	\end{align*}

So we can finally write the connected four-point Green's function up to one loop order
 	\begin{align*}
	 \tilde{G}_{c;2}^{(4)}(k_1,...,&k_4)   =\sum_{j=1}^6 \tilde{G}_{\begin{NoHyper}\ref{fig3}\end{NoHyper}\text{-}j}^{(4)} +  \sum_{n=1}^3\sum_{m=1}^{18}\tilde{G}_{\begin{NoHyper}\ref{fig5}\end{NoHyper}\text{-}nm}^{(4)} +\sum_{i=1}^4\sum_{j=1}^6\tilde{G}_{\begin{NoHyper}\ref{fig7}\end{NoHyper}\text{-}ij}^{(4)} +  \sum_{i=1}^4\sum_{j=1}^6\tilde{G}_{\begin{NoHyper}\ref{fig8}\end{NoHyper}\text{-}ij}^{(4)}\\ 
	=-\Bigg(& \sum_j  \frac{ig}{6} \frac{ e^{-\sum_a \bar \eta(k_a)-\frac{i\theta}{2} E_j}}{\prod_a ( k_a^2 - m^2)} \\
	+& \sum_{nm}\frac{g^2}{36} \int \frac{d^3 q}{(2\pi)^3} \frac{ e^{-\sum_a \bar\eta(k_a)-i\frac{\theta}{2} F_{nm}} }{(q^2-m^2)((p_n-q)^2-m^2)\prod_a ( k_a^2 - m^2)}  \\
	+& \sum_{ij}\frac{g^2}{36} \int \frac{d^3 q}{(2\pi)^3} \frac{\left( 2+ e^{-i \theta \boldsymbol k_i \wedge \boldsymbol q} \right) \, e^{-\sum_a \bar \eta(k_a)-i\frac{\theta}{2} E_j }  \;  }{(q^2-m^2)(k_i^2-m^2)\prod_a ( k_{a}^2 - m^2)} \Bigg) \delta^{(3)}\left( \sum k_a \right)
	\end{align*}
As we said in the last section of chapter \ref{chap-s-ord}, the ultraviolet/infrared mixing is present in the Green's functions of all the s-ordered products. We noted that this was due to the fact that it is the Moyal part the responsible of the mixing. From the results of this chapter it is clear that the mixing is present for any translation invariant star product, and it is again the Moyal part the only one that contributes to the mixing. This is consistent with the fact that the Moyal part is the only responsible of the noncommutativity of any translation invariant star product.

\chapter{Translation invariant star products as twisted products} \label{chaptransinv}

In the last two chapters we considered the noncommutative $(2+1)$-dimensional space-time with two noncommutative spacial coordinates and one time coordinate which commutes with the two spacial coordinates. More precisely, considered the noncommutative space-time with the following commutation relations
 	\begin{equation}
	[x^{\mu},x^{\nu}]_{\star}=i \theta^{\mu \nu} \label{53}
	\end{equation}
with $\mu,\nu \in \{ 1,2,3\}$, and 
	\begin{equation*}
	 \left( \theta^{\mu \nu}\right) = \left(
		\begin{matrix}
		0 & 1 & 0\\
		-1 & 0 & 0\\
		0 & 0 & 0
		\end{matrix} \right)
	\end{equation*}
where the $\star$ refers either to the s-ordered product (studied in chapter \ref{chap-s-ord}) or more generally to any translation invariant star product (studied in chapter \ref{chap-transinv}). The $\theta^{\mu \nu}$ is a parameter of the noncommutativity of space. The commutator \eqref{53} is invariant under coordinate translations, but it is not invariant under Lorentz transformations \cite{chaichian}, which means that $\theta^{\mu \nu}$ is not invariant under Lorentz transformations. This is not good for a fundamental theory because the noncommutativity should be an intrinsic property of space and not a frame dependent property. We will see that this is not as bad as it may seem, because the Poincar\'e invariance of the $\theta^{\mu \nu}$ parameter is satisfied at a deformed level\footnote{This has already been shown for quantum field theory in the Moyal case \cite{chaichian}}, as we will explain in this chapter.

\section{Differential form of a general translation invariant star product}

In this section we derive the differential form of a general translation invariant star product, which will be useful later. Recall that a general translation invariant star product was given by \eqref{36}
 	\begin{equation*}
	(f \star g)(x)= \int \frac{d^2p}{(2\pi)^2} \frac{d^2q}{(2\pi)^2} \tilde f(q) \tilde g(p-q) e^{\alpha (p,q)} e^{-ip \cdot x} 
	\end{equation*}
or equivalently
 	\begin{equation}
	(f \star g)(x)= \int \frac{d^2p}{(2\pi)^2} \frac{d^2q}{(2\pi)^2} \tilde f(q) \tilde g(p) e^{\alpha (p+q,q)} e^{-i(p+q) \cdot x}. \label{60}
	\end{equation}
We assume that the $\alpha$ function can be expanded as
 	\begin{equation*}
	\alpha (p,q) = \sum_{\vec i, \, \vec j}\alpha_{\vec i,  \vec j} \, \boldsymbol p^{\vec i}  \boldsymbol q^{\vec j} 
	\end{equation*}
for $\alpha_{\vec i,  \vec j} \in \mathbb C$, where we use the notation $\boldsymbol p^{\vec i} = p^{i_1}p^{i_2}$ and the sum goes from zero to infinity. With this notation, equation \eqref{60} becomes
 	\begin{equation}
	(f \star g)(x)= \int \frac{d^2p}{(2\pi)^2} \frac{d^2q}{(2\pi)^2} \tilde f(q) \tilde g(p) \text{exp}\left(  \sum_{\vec i, \, \vec j}\alpha_{\vec i,  \vec j} \, (\boldsymbol {p+q})^{\vec i}  \boldsymbol q^{\vec j}  \right) e^{-i(p+q) \cdot x}
	\end{equation}
but we can easily see that the $p's$ and $q's$ in the expansion of the $\alpha$ function can be substituted by partial derivatives acting on the exponential. More precisely, the expression can be written as
 	\begin{align*}
	(f &\star g)(x) =\\& \int \frac{d^2p}{(2\pi)^2} \frac{d^2q}{(2\pi)^2} \tilde f(q) \tilde g(p)  e^{-iq \cdot x} \text{exp}\left(  \sum_{\vec i, \, \vec j}\alpha_{\vec i,  \vec j} \, (i\overrightarrow \partial_x+i\overleftarrow \partial_x)^{\vec i}  (i\overleftarrow \partial_x)^{\vec j}  \right) e^{-ip \cdot x}
	\end{align*}
where $\overrightarrow \partial_x^{\vec i}=\overrightarrow \partial_{x_1}^{ i_1}\overrightarrow \partial_{x_2}^{i_2}$ acts on the exponential to the right and  $\overleftarrow \partial_x^{\vec i}=\overleftarrow \partial_{x_1}^{ i_1}\overleftarrow \partial_{x_2}^{i_2}$ acts on the exponential to the left. Performing the integrals, we can write this as a series expansion 
 	\begin{align*}
	(f \star g)(x) =  f(x)  \text{exp}\left(  \sum_{\vec i, \, \vec j}\alpha_{\vec i,  \vec j} \, (i\overrightarrow \partial_x+i\overleftarrow \partial_x)^{\vec i}  (i\overleftarrow \partial_x)^{\vec j}  \right) g(x)
	\end{align*}
which we write as 
 	\begin{align}
	(f \star g)(x) =  f(x)  e^{\alpha (i\overrightarrow \partial_x+i\overleftarrow \partial_x,  i\overleftarrow \partial_x)} g(x). \label{64}
	\end{align}
Is is important to note that the differential expression may be valid on a smaller range (depending on the $\alpha$ function) than the integral expression, so from now on we assume that the functions are in the range of the differential expression of the star product.

\section{Twisted Poincar\'e algebra}

We start by recalling that the Poincar\'e algebra is characterized by the Lorenz generators $M_{\mu \nu}$ and the translations generators $P_{\mu}$, whose representation on the algebra of functions on Minkowski space-time are given by 
 	\begin{align*}
		 P_{\mu} & = -i\partial_{\mu} \\
		M_{\mu \nu}&= i(x_{\mu} \partial_{\nu}-x_{\nu}\partial_{\mu})
	\end{align*}
and satisfy the following commutation relations
 	\begin{align}
		[M_{\mu \nu},M_{\rho \sigma}]&= i(\eta_{\mu \sigma}M_{\nu \rho}-\eta_{\nu \sigma}M_{\mu \rho}+\eta_{\nu \rho}M_{\mu \sigma}-\eta_{\mu \rho}M_{\nu \sigma})\nonumber \\
		[M_{\mu \nu}, P_{\rho}] & = i(\eta_{\mu \rho}P_{\nu}-\eta_{\nu \rho}P_{\mu}) \label{78} \\ 
		[P_{\mu},P_{\nu}]&=0 \nonumber
	\end{align}
Moreover, its universal enveloping algebra $U(\mathcal P)$ has a noncommutative, but cocommutative Hopf algebra structure \cite{aschieri1} with coproduct, counit and antipode given respectively by
 	\begin{align}
		\Delta_0 (X)&= X\otimes 1 +1 \otimes X,  &\Delta_0 (1)=1 \otimes 1 \label{68}\\
		\varepsilon_0 (X) & =0,  &\varepsilon(1)=1\label{75} \\
		S_0(X)&=-X,  &S(1) = 1
	\end{align}
where $X$ stands for $M_{\mu \nu}$ and $P_{\mu}$. Let us now define the twist $\mathcal F$ as an invertible element of $U(\mathcal P) \otimes U(\mathcal P)$ \cite{twist} such that
 	\begin{align}
		\mathcal F_{12}(\Delta \otimes id) \mathcal F&= \mathcal F_{23}(id \otimes \Delta) \mathcal F \label{56} \\
		(\varepsilon \otimes id) \mathcal F & = (id \otimes \varepsilon) \mathcal F  = 1 \label{76}
	\end{align}
where
 	\begin{align*}
		 \mathcal F_{12}= \mathcal F \otimes 1 \quad \text{and} \quad  \mathcal F_{23}= 1 \otimes  \mathcal F.
	\end{align*}
We will sometimes write the twist and its inverse as (sum over $ \alpha$ understood)
 	\begin{equation*}
		 \mathcal F = \mathrm f^{\alpha} \otimes \mathrm f_{\alpha}\quad \text{and} \quad  \mathcal F^{-1} =  \bar {\mathrm f}^{\alpha} \otimes \bar {\mathrm f}_{\alpha}.
	\end{equation*}
Recalling that the ordinary product between functions on the space-time is defined as
 	\begin{align*}
		 m_0 : \text{Fun}(\mathcal M) \otimes  \text{Fun}(\mathcal M) &\to  \text{Fun}(\mathcal M) \\
		h \otimes g &\mapsto hg
 	\end{align*}
we define the twisted product of functions as
 	\begin{equation}
		m_{\star}(h \otimes g) = m_0 \circ  \mathcal F^{-1} (h \otimes g)=  \bar {\mathrm f}^{\alpha}(h) \bar {\mathrm f}_{\alpha}(g) \label{65}
	\end{equation}
regarding the twist $ \mathcal F$ as a map 
	\begin{equation*}
		\text{Fun}(\mathcal M) \otimes \text{Fun}(\mathcal M) \to \text{Fun}(\mathcal M) \otimes \text{Fun}(\mathcal M).
	\end{equation*}
The condition \eqref{56} guarantees that the twisted product of functions is associative, and  \eqref{76} is equivalent to conditions \eqref{73} and \eqref{74}, i.e. it says that the multiplication by a constant function is just ordinary scalar multiplication.  It is now easy to see that the s-ordered product can be written in terms of a twist as
 	\begin{equation*}
		h \star_s g  = m_0 \circ  \mathcal F_s^{-1} (h \otimes g)  
	\end{equation*}
where
 	\begin{equation*}
		\mathcal F_s^{-1} =e^{\frac{\theta}{2}[s(\partial_1 \otimes \partial_1 + \partial_2 \otimes \partial_2)+i(\partial_1 \otimes \partial_2-\partial_2 \otimes \partial_1)]} 
	\end{equation*}
In fact, it is possible to write any translation invariant star product in terms of a twist. Indeed, from equation \eqref{64} we have 
 	\begin{align*}
		(f \star g)(x) &=  f(x)  e^{\alpha (i\overrightarrow \partial_x+i\overleftarrow \partial_x,  i\overleftarrow \partial_x)} g(x) \nonumber \\
		&= m_0 \circ e^{\alpha (1 \otimes i \partial_x+i \partial_x \otimes 1,  i \partial_x \otimes 1)}(f(x) \otimes g(x))
	\end{align*}
Comparing with equation \eqref{65} we can see that the corresponding twist is given by
 	\begin{equation}
		\mathcal F^{-1} = e^{\alpha (1 \otimes i \partial+i \partial \otimes 1,  i \partial \otimes 1)} \label{66}
	\end{equation}
which, using equation \eqref{37} can be written as 
 	\begin{align}
		\mathcal F^{-1} &= e^{\alpha (1 \otimes i \partial+i \partial \otimes 1,  i \partial \otimes 1)} \nonumber \\
		&= e^{\eta (i \partial \otimes 1) - \eta (1 \otimes i \partial+i \partial \otimes 1) + \eta (1 \otimes i \partial) + i\omega(1 \otimes i \partial,i \partial \otimes 1)} \nonumber \\
		&= e^{\eta (i \partial) \otimes 1 + 1 \otimes \eta( i \partial)- \eta (1 \otimes i \partial+i \partial \otimes 1)  - \frac{i \theta}{2}(i\partial_1 \otimes i\partial_2 - i\partial_2 \otimes i\partial_1)} \label{71}
	\end{align}
where $\eta (i \partial)=\eta (i \partial_1,i \partial_2)$. Written in terms of the generators $P_{\mu}=-i\partial_{\mu}$, the twist is
 	\begin{align}
		\mathcal F^{-1} = e^{\bar\eta (P) \otimes 1 + 1 \otimes \bar\eta(P)- \bar\eta (1 \otimes P+P \otimes 1)  - \frac{i \theta}{2}(P_1 \otimes P_2 - P_2 \otimes P_1)}.
	\end{align}
 where we used the fact that $\eta(-x)=\bar \eta(x)$. We still have to check conditions \eqref{56} and \eqref{76}. As we said, the cocycle condition on the twist (equation \eqref{56}), is just the associativity condition of the star product. Indeed, from the expression \eqref{66} we have  
 	\begin{equation}
		\mathcal F_{12} = e^{-\alpha(i\partial \otimes 1 \otimes 1 + 1 \otimes i\partial  \otimes 1, i\partial  \otimes 1  \otimes 1)}
	\end{equation}
and
 	\begin{equation}
		\mathcal F_{23} = e^{-\alpha(1\otimes i\partial \otimes 1  + 1\otimes1 \otimes i\partial, 1\otimes i\partial  \otimes 1 )}
	\end{equation}
while using equation \eqref{68} and the fact that $\Delta(ab) = \Delta(a)\Delta(b)$, we have
 	\begin{equation}
		(\Delta \otimes id) \mathcal F = e^{-\alpha(i\partial \otimes 1 \otimes 1 +1\otimes 1 \otimes i\partial +1\otimes  i \partial \otimes 1, i\partial  \otimes 1  \otimes 1 + 1\otimes i \partial \otimes 1)}
	\end{equation}
and
 	\begin{equation}
		(id \otimes \Delta) \mathcal F = e^{-\alpha(i\partial \otimes 1 \otimes 1 +1\otimes i\partial \otimes 1+1\otimes 1\otimes  i \partial , i\partial  \otimes 1  \otimes 1)}
	\end{equation}
But $\partial \otimes 1 \otimes 1$, $1\otimes i\partial \otimes 1$ and $1\otimes 1 \otimes i\partial $ must be taken as independent variables, so equation \eqref{56} says 
 	\begin{align*}
		\alpha(r+ q, r)+ \alpha(r+s +q, r + q) = \alpha(q  + s, q) + \alpha(r+q+s , r)
	\end{align*}
where we used the correspondence $\partial \otimes 1 \otimes 1 \to r$, $1\otimes i\partial \otimes 1 \to q$ and $1\otimes 1 \otimes i\partial \to s$. But setting $r+q+s \to p$ and $q \to q-r$ we get
 	\begin{align*}
		\alpha(q, r)+ \alpha(p, q) = \alpha(p-r, q-r) + \alpha(p , r)
	\end{align*}
which is the associativity condition \eqref{70}. This condition is already satisfied by the $\alpha$ function, so the twist \eqref{71} satisfies the cocycle condition \eqref{56}. On the other hand, equation\eqref{76} is equivalent to equations \eqref{73} and \eqref{74}. Indeed, using equations \eqref{75} and the fact that $\varepsilon(ab)=\varepsilon(a)\varepsilon(b)$ we have 
 	\begin{align*}
		(\varepsilon \otimes id) \mathcal F =(\varepsilon \otimes id) e^{-\alpha (1 \otimes i \partial_x+i \partial_x \otimes 1,  i \partial_x \otimes 1)} =e^{-\alpha (1 \otimes i \partial_x,0) }
	\end{align*}
and
 	\begin{align*}
		 (id \otimes \varepsilon) \mathcal F = (id \otimes \varepsilon)e^{-\alpha (1 \otimes i \partial_x+i \partial_x \otimes 1,  i \partial_x \otimes 1)} =e^{-\alpha (i \partial_x \otimes 1,i \partial_x \otimes 1) }
	\end{align*}
So equation \eqref{76} is equivalent to equations \eqref{73} and \eqref{74}, which are already satisfied by the $\alpha$ function. This means that the expression \eqref{71} is indeed a twist.  

In the previous chapter we studied the translation invariant star products, i.e. those that satisfy the following condition  
 	\begin{equation}
	\mathcal T_a (h \star g)= \mathcal T_a(h) \star \mathcal T_a(g) \label{61}
	\end{equation}
where $\mathcal T_a (h)(x) = h(x+a)$. The infinitesimal form of this condition is the Leibniz rule
 	\begin{equation*}
		P_{\mu}(h \star g)=(P_{\mu}h) \star g+h \star (P_{\mu}\, g)
	\end{equation*}
We know that Quantum Field Theory, besides being translation invariant, is also Lorentz invariant. The generators of the Poincar\'e algebra are derivations of the pointwise product, that is
 	\begin{align}
		P_{\mu}(h  g)&=(P_{\mu}h)  g+h  (P_{\mu}\, g) \nonumber \\
		M_{\mu \nu}(h g)&=(M_{\mu \nu}h) g+h (M_{\mu \nu} g). \label{62}
	\end{align}
However, for the star products only the first of these two equations is satisfied while the second one is not. In Noncommutative Quantum Field Theory the translation invariance is preserved but the Lorentz invariance is broken \cite{chaichian}. This is reflected in the fact that the noncommutative parameter $\theta_{\rho \sigma}$ is not invariant under Lorentz transformations. As we said, this is not satisfying for a fundamental theory, but we will see that with the help of the twist there is a way to achieve the Poincar\'e invariance of the noncommutative parameter $\theta_{\rho \sigma}$ at a deformed level. To see this, first note that the Leibniz rule, i.e. the action of the Poincar\'e generators $X$ on the ordinary product of functions, can be written in terms of the product and coproduct as follows
 	\begin{equation}
		X(hg)=(Xh)g+h(Xg)=m_0(\Delta (X)(h \otimes g)) \label{57}
	\end{equation}
So the action on the star product of functions is given by
 	\begin{align}
		X(h \star g)&=X(\bar {\mathrm f}^{\alpha}(h) \bar {\mathrm f}_{\alpha}(g)) \nonumber \\
		&=m_0 \circ \Delta (X)( \bar {\mathrm f}^{\alpha}(h)  \otimes \bar {\mathrm f}_{\alpha}(g) ) \nonumber \\
		&=m_0 \circ \Delta (X) \mathcal F^{-1} ( h \otimes g ) \nonumber \\
		&= m_{\star} \circ \mathcal F \Delta (X) \mathcal F^{-1} ( h \otimes g )
	\end{align}
where we used the fact that the product between $\bar {\mathrm f}^{\alpha}(h)$ and $\bar {\mathrm f}_{\alpha}(g)$ is the ordinary product. Comparing the last two equations we can easily see that it is possible to have a deformed version of the Leibniz rule if we define the twisted coproduct as 
 	\begin{equation}
		\Delta_{\mathcal F} (X)= \mathcal F \Delta (X) \mathcal F^{-1}. \label{72}
	\end{equation}
So the twisted action of the Poincar\'e generators on the star product of functions is given by
 	\begin{equation}
		X(h \star g) = m_{ \star} \circ \Delta_{\mathcal F} (X) ( h \otimes g ). \label{58}
	\end{equation}
The twisted universal enveloping algebra $U_{\mathcal F}(\mathcal P)$ is then defined to be the algebra generated by $P_{\mu}$ and $M_{\mu \nu}$ modulo the commutation relations \eqref{78} with coproduct given by equation \eqref{72} and counit and antipode as in the undeformed case. Note that the fact that the commutation relations are unchanged means that we will have the same representations of the ordinary Poincar\'e algebra. This is not the only way of twisting the universal enveloping algebra in this framework, another approach, in which the generators and the commutator are deformed, has also been considered \cite{lizzivitaletwistall}.   

The symmetries of the noncommutative parameter $\theta_{\rho \sigma}$ are given by the poincar\'e generators acting with the rule \eqref{58}. To see this we will need the explicit expression of equation  \eqref{72} for the general translation invariant star product, which we compute with the help of the following operator formula:
 	\begin{equation}
		Ad\, e^B (C)=e^B C e^{-B}= \sum_{n=0}^{\infty}\frac{1}{n!}[B,[B,...[B,C]=\sum_{n=0}^{\infty}\frac{(AdB)^n}{n!}C. \label{85}
	\end{equation}
In our case we have $C=\Delta (X)=X \otimes 1 + 1 \otimes X$ and $B=-\bar\eta (P) \otimes 1 -1 \otimes \bar\eta(P)+ \bar\eta (1 \otimes P+P \otimes 1)  + \frac{i \theta}{2}(P_1 \otimes P_2 - P_2 \otimes P_1)$. Given that the momentum operators commute between them, it is clear that we have 
 	\begin{equation}
		\Delta_{\mathcal F} (P_{\mu}) = \Delta (P_{\mu}).
	\end{equation}
The action of $P_{\mu}$ on the commutator $[x_{\rho},x_{\sigma}]_{\star}$ is then given by 
 	\begin{align*}
		P_{\mu}([x_{\rho},x_{\sigma}]_{\star}) &= m_{ \star} \circ \Delta (P_{\mu}) ( x_{\rho}\otimes x_{\sigma} -x_{\sigma}\otimes x_{\rho})\\
		&= m_{ \star}(-i\delta_{\mu \rho} \otimes x_{\sigma}-x_{\rho} \otimes i\delta_{\mu \sigma}+i\delta_{\mu \sigma} \otimes x_{\rho}+x_{\sigma} \otimes i\delta_{\mu \rho})=0
	\end{align*}
So the twisted action of the translation generators are symmetries of the $\theta_{\rho \sigma}$ parameter. This was clearly expected given that the action of the translation generators was not deformed and they where already symmetries of the space. On the other hand, the twisted action of the Lorentz generators on the commutator is a bit more involved. This is computed in appendix \ref{sec:twistedgenerators} and is found to be zero
 	\begin{align*}
		 M_{\mu \nu}([x_{\rho}, x_{\sigma}]_\star)=0
	\end{align*}
This result, which is actually valid in any dimension, and the fact that the commutator is equal to $i\theta_{\rho \sigma}$, indicates that the parameter $\theta_{\rho \sigma}$ is invariant under the action of the twisted Poincar\'e generators, which is compatible with $\theta$ being constant.

\chapter*{Conclusions and Outlook}\addcontentsline{toc}{chapter}{Conclusions and Outlook}

In this work we have considered the $\phi^{\star 4}$ field theory for the s-ordered products and then for the general translation invariant star products. We have computed the propagator and the vertex of the theory. Given that the vertex is not invariant under arbitrary permutations, we computed all the non-equivalent diagrams of the two-point and four-point Green's functions up to one loop and their corresponding integrals.  We found that the diagrams and symmetry factors are the same for all translation invariant star products at any order.  

In the last chapter we derived the differential expression of a general translation invariant star product and we then found that any product of this type can be written as a twist. We found that the noncommutative parameter $\theta^{\rho \sigma}$ is invariant under the twisted action of the Poincar\'e generators, which means that for any translation invariant star product, $\theta^{\rho \sigma}$ is Poincar\'e invariant at a deformed level.

Given that the Green's functions depend on the particular star product being used (they depend on the $\eta$ function), a possible continuation to this work is the computation of the S matrix of the theory, to compare the different star products at the level of physical quantities. This is motivated by the fact that there are two point of views: on one hand, one can think that the physical quantities should just depend on the noncommutative structure of space (i.e. the commutator $[x^i,x^j]_{\star} = i \theta^{ij}$) not in the specific product used for the computations. On the other hand one may think that the space is the ordinary one, but the action is different and contains an infinite number of derivatives, in which case one would expect different physical results, as it has already been argued in \cite{lizzivitaletwist}. We can also mention another point of view: when one considers ``noncommutative spaces", given that the notion of points is lost and strictly speaking one does not have a space anymore, the new playground (or new ``space") is the noncommutative algebra defined by the star product, in which case the physical quantities may depend on the particular product.\footnote{Thank you to Professor Thierry Masson for clarifying this point.}

\chapter*{Acknowledgments}\addcontentsline{toc}{chapter}{Acknowledgments}

This work was done in Universit\`a degli Studi di Napoli Federico II and corresponds to my thesis of the master ``Physique Th\'eorique et Math\'ematique, Physique des Particules et Astrophysique" of Aix-Marseille University. I am very grateful to my supervisors Fedele Lizzi and Patrizia Vitale for giving me the oportunity of doing this internship with them and for guiding me through this wonderful world of noncommutative geometry; this project wouldn't have been possible without their help. I also want to thank Thierry Masson; thanks to him I discovered the field of noncommutative geometry. He introduced me into this field and helped me doing the first steps towards this project. As a student of the master ``Physique Th\'eorique et Math\'ematique, Physique des Particules et Astrophysique" in Aix-Marseille universit\'e, I want to thank Alejandro Perez and Arnaud Duperrin who are in change of the master, and to the university for funding my internship project. And last, but not least, I want to thank my family and my girlfriend who have always supported me and encouraged me throughout my studies.    
 
\newpage


\appendix
\addtocontents{toc}{\protect\newpage}
\chapter{Cyclic invariance of the vertex}
\label{sec:app}

	\begin{property}
	If $\sum_{c=1}^n k_c = 0$, then, the function 
		\begin{equation*}
		F(k_1,...,k_n) = \sum_{a<b}  \, k_a \wedge k_b
		\end{equation*}
	 is invariant under cyclic permutations.
	\end{property}

	\begin{proof}
	We want to prove that $F(k_1,...,k_n) = F(\sigma (k_1),...,\sigma(k_n))$ where $\sigma(a)=(a+k)\, mod\,n$, with $0 \leq k < n$.\footnote{Here we take the function mod as 
\begin{equation*}
 a\,mod\,n =   
	\begin{cases}
	a+k & \text{if } a + k \leq n,\\
	a+k-n & \text{if } a + k > n
	\end{cases}
\end{equation*}

where  $0 \leq k < n$.} Beginning from the right hand side of the equation we have

	\begin{equation*}
	F(\sigma (k_1),...,\sigma(k_n)) = \sum_{a<b} \,  k_{\sigma(a)} \wedge k_{\sigma (b)}
	\end{equation*}

	\begin{eqnarray*}
	= \sum_{0<a<b\leq n-k}\quad \sum_{n-k<a<b \leq n} \quad \sum_{0<a\leq n-k < b \leq n} \,  k_{\sigma(a)} \wedge k_{\sigma (b)}
	\end{eqnarray*}

	\begin{eqnarray*}
	= \sum_{0<a<b\leq n-k}\,  k_{a+k} \wedge k_{b+k} + \sum_{n-k<a<b \leq n} \,  k_{a+k-n} \wedge k_{b+k-n}  + \sum_{0<a\leq n-k < b \leq n} \,  k_{a+k} \wedge k_{b+k-n}
	\end{eqnarray*}

	\begin{eqnarray*}
	= \sum_{0<a-k<b-k\leq n-k} \,  k_a \wedge k_b + \sum_{n-k<a-k+n<b-k+n \leq n} \,  k_a \wedge k_b \\ + \sum_{0<a-k\leq n-k < b-k+n \leq n} \,  k_a \wedge k_b 
	\end{eqnarray*}

	\begin{eqnarray*}
	= \sum_{k<a<b\leq n} \,  k_a \wedge k_b + \sum_{0<a<b \leq k} \,  k_a \wedge k_b  + \sum_{0<b\leq k < a \leq n} \,  k_a \wedge k_b
	\end{eqnarray*}
but we can exchange the dummy indices in the last term, and use the fact that  $k_a \wedge k_ b$ is antisymmetric, so we get

	\begin{eqnarray*}
	= \sum_{k<a<b\leq n} \, k_a \wedge k_b + \sum_{0<a<b \leq k}  \,  k_a \wedge k_b - \sum_{0<a\leq k < b \leq n} \,  k_a \wedge k_b
	\end{eqnarray*}

	\begin{eqnarray*}
	= \sum_{a<b} \, k_a \wedge k_b - 2\sum_{0<a\leq k < b \leq n} \,  k_a \wedge k_b
	\end{eqnarray*}

	\begin{eqnarray*}
	= F(k_1,...,k_n)  - 2\sum_{0<a\leq k < b \leq n} \,  k_a \wedge k_b.
	\end{eqnarray*}
So we just have to prove that the term $\sum_{0<a\leq k < b \leq n} \,  k_a \wedge k_b$ is zero. Using the fact that  $k_n = -\sum_{0<c<n-1} k_c $, we have that

	\begin{equation*}
	\sum_{0<a\leq k < b \leq n} \,  k_a \wedge k_b = \sum_{0<a\leq k < b < n} \,  k_a \wedge k_b + \sum_{0<a\leq k} \,  k_a \wedge k_n
	\end{equation*}

	\begin{equation*}
	= \sum_{0<a\leq k < b < n} \,  k_a \wedge k_b - \sum_{\substack{0<a\leq k \\ 0<c<n-1}} \,  k_a \wedge k_c
	\end{equation*}

	\begin{equation*}
	= \sum_{0<a\leq k < b < n} \,  k_a \wedge k_b - \sum_{0<a\leq k <c<n-1} \,  k_a \wedge k_c - \sum_{\substack{0<a\leq k \\ 0<c \leq k}} \,  k_a \wedge k_c
	\end{equation*}

	\begin{equation*}
	=  - \sum_{\substack{0<a\leq k \\ 0<c \leq k}} \,  k_a \wedge k_c = 0
	\end{equation*}
where the last equality follows from the fact that $k_a \wedge k_c$ is antisymmetric.
	\end{proof}

\chapter{Properties of the displacement operator}
\label{sec:appB}

In this appendix we show some of the  properties used in section \ref{sec:s-ord}. We begin with the following property

	\begin{equation}
	W(\omega) W(\omega ') = e^{\frac{i}{\theta} \text{Im}(\omega \bar {\omega}')}W(\omega + \omega ') \label{b3}
	\end{equation}
 where $W(\omega)= e^{\frac{1}{\theta}(\omega \hat{a}^\dag-\bar{\omega}\hat{a})}$. First, note that using the Baker-Campbell-Hausdorff formula, we have the following two relations

	\begin{equation}
	e^{\frac{1}{\theta}(\omega' \hat{a}^\dag - \bar{\omega} \hat{a})} = e^{\frac{1}{\theta}\omega'  \hat{a}^\dag} e^{-\frac{1}{\theta} \bar{\omega} \hat{a}}e^{-\frac{1}{2\theta}\omega' \bar{\omega}} \quad \text{and} \quad e^{\frac{1}{\theta}(\omega' \hat{a}^\dag - \bar{\omega} \hat{a})} = e^{-\frac{1}{\theta} \bar{\omega} \hat{a}}  e^{\frac{1}{\theta}\omega'  \hat{a}^\dag} e^{\frac{1}{2\theta}\omega' \bar{\omega}} \label{b2}
	\end{equation} 
which imply

	\begin{equation}
	 e^{\frac{1}{\theta}\omega'  \hat{a}^\dag} e^{-\frac{1}{\theta} \bar{\omega} \hat{a}}  = e^{-\frac{1}{\theta} \bar{\omega} \hat{a}}  e^{\frac{1}{\theta}\omega'  \hat{a}^\dag} e^{\frac{1}{\theta}\omega' \bar{\omega}} \label{b1}
	\end{equation}
So the left hand side of equation \eqref{b3} is

	\begin{equation*}
	W(\omega) W(\omega ') =  e^{\frac{1}{\theta}\omega  \hat{a}^\dag}  e^{-\frac{1}{\theta} \bar{\omega} \hat{a}}  e^{\frac{1}{\theta}\omega'  \hat{a}^\dag} e^{-\frac{1}{\theta} \bar{\omega'} \hat{a}}  e^{-\frac{1}{2\theta}(\vert \omega \vert^2 + \vert \omega' \vert^2)}
	\end{equation*}
While the right hand side is

	\begin{equation*}
	W(\omega + \omega ') = e^{\frac{1}{\theta}(\omega +\omega')  \hat{a}^\dag} e^{-\frac{1}{\theta} (\bar{\omega}+\bar{\omega}') \hat{a}}e^{-\frac{1}{2\theta}\vert \omega + \omega' \vert^2} = e^{\frac{1}{\theta}\omega  \hat{a}^\dag} e^{\frac{1}{\theta}\omega'  \hat{a}^\dag} e^{-\frac{1}{\theta} \bar{\omega} \hat{a}}  e^{-\frac{1}{\theta} \bar{\omega'} \hat{a}}  e^{-\frac{1}{2\theta}(\vert \omega \vert^2 + \vert \omega' \vert^2 +\omega \bar \omega' + \bar \omega \omega')}
	\end{equation*}

	\begin{equation*}
	= e^{\frac{1}{\theta}(\omega +\omega')  \hat{a}^\dag} e^{-\frac{1}{\theta} (\bar{\omega}+\bar{\omega}') \hat{a}}e^{-\frac{1}{\theta}\frac{1}{2}\vert \omega + \omega' \vert^2} = e^{\frac{1}{\theta}\omega  \hat{a}^\dag} e^{\frac{1}{\theta}\omega'  \hat{a}^\dag} e^{-\frac{1}{\theta} \bar{\omega} \hat{a}}  e^{-\frac{1}{\theta} \bar{\omega'} \hat{a}}  e^{-\frac{1}{2\theta}(\vert \omega \vert^2 + \vert \omega' \vert^2 +\omega \bar \omega' + \bar \omega \omega')}
	\end{equation*}

but using equation \eqref{b1} this is

	\begin{equation*}
	 = e^{\frac{1}{\theta}\omega  \hat{a}^\dag} e^{-\frac{1}{\theta} \bar{\omega} \hat{a}} e^{\frac{1}{\theta}\omega'  \hat{a}^\dag}  e^{-\frac{1}{\theta} \bar{\omega'} \hat{a}}  e^{-\frac{1}{2\theta}(\vert \omega \vert^2 + \vert \omega' \vert^2 +\omega \bar \omega' + \bar \omega \omega')}e^{\frac{1}{\theta}\omega' \bar{\omega}} = W(\omega) W(\omega ')  e^{-\frac{1}{2\theta}(\omega \bar \omega' - \bar \omega \omega')}
	\end{equation*}
So finally we have

	\begin{equation*}
	W(\omega + \omega ') = W(\omega) W(\omega ')  e^{-\frac{i}{\theta} \text{Im}(\omega \bar \omega)}
	\end{equation*}
which is what we wanted. Now let's prove the following property

	\begin{equation*}
	\text{Tr} \, W(\omega) = \pi\theta \delta^{(2)}(\omega)
	\end{equation*}
Beginning from the left hand side, using relations \eqref{b2}, and using the complete set of coherent states $\ket{\alpha}=W(\alpha)\ket{0}$ to compute the trace (see \cite{coherent}) note that our expressions differ from those of the reference by a factor of $\theta$ due to our conventions of $\hat a$ and $\alpha$), we have

	\begin{equation*}
	\text{Tr} \, W(\omega) = \text{Tr} \, \left(  e^{\frac{1}{\theta}(\omega \hat{a}^\dag-\bar{\omega}\hat{a})} \right) = \int d^2 \alpha \,(\pi\theta)^{-1} \bra{\alpha}e^{\frac{1}{\theta}(\omega \hat{a}^\dag-\bar{\omega}\hat{a})} \ket{\alpha}
	\end{equation*}

	\begin{equation*}
	= \int d^2 \alpha \,(\pi\theta)^{-1}  \bra{\alpha} e^{\frac{1}{\theta}\omega  \hat{a}^\dag} e^{-\frac{1}{\theta} \bar{\omega} \hat{a}}e^{-\frac{1}{2\theta}\omega \bar{\omega}} \ket{\alpha} = \int d^2 \alpha \,(\pi\theta)^{-1}  e^{\frac{1}{\theta}\omega  \bar{\alpha}} e^{-\frac{1}{\theta} \bar{\omega} \alpha}e^{-\frac{1}{2\theta}\omega \bar{\omega}} 
	\end{equation*}

	\begin{equation*}
	= e^{-\frac{1}{2\theta}\omega \bar{\omega}} \,(\pi\theta)^{-1} \int d^2 \alpha \;  e^{\frac{2i}{\theta}(\omega_2 \alpha_1 - \omega_1 \alpha_2)}=e^{-\frac{1}{2}\omega \bar{\omega}} \frac{4\pi}{\theta} \delta^{(2)}(2\omega/\theta) = \pi\theta \delta^{(2)}(\omega) 
	\end{equation*}

\chapter{Twisted action of the Lorentz generators}
\label{sec:twistedgenerators}

In this appendix we compute the explicit expression of $\Delta_{\mathcal F} (M_{\mu \nu})$, and with the help of that expression we will compute the twisted action of $M_{\mu \nu}$ on the commutator  $[x_{\rho},x_{\sigma}]_{\star}$. 

Note that the commutator $[P_{\alpha},M_{\mu \nu}]$   commutes with $P_{\alpha}$, so we can use the following well known property
 	\begin{equation*}
		[f(P_{\alpha}),M_{\mu \nu}]=f'(P_{\alpha})[P_{\alpha},M_{\mu \nu}]
	\end{equation*}
Using this we have\footnote{Note that we abuse of notation. We omit the hat in the operators $\hat P_{\mu}$, and by $dP_{\mu}^n/dP_{\mu}$ we mean $dx^n/dx \vert_{x=\hat P_{\mu}}$} 
 	\begin{align*}
		[P_1^n P_2^m,M_{\mu \nu}]&=P_1^n[ P_2^m,M_{\mu \nu}]+[P_1^n ,M_{\mu \nu}]P_2^m \\
		&=P_1^n\frac{d}{dP_2} P_2^m[ P_2,M_{\mu \nu}]+[P_1 ,M_{\mu \nu}]\frac{d}{dP_1} P_1^n P_2^m \\
		&=\frac{\partial}{\partial P_2} (P_1^n P_2^m)[ P_2,M_{\mu \nu}]+\frac{\partial}{\partial P_1} (P_1^n P_2^m)[P_1 ,M_{\mu \nu}]
	\end{align*}
or in general we have
 	\begin{align*}
		[\eta(P_1, P_2),M_{\mu \nu}]=\partial_2 \eta [ P_2,M_{\mu \nu}]+\partial_1 \eta [P_1 ,M_{\mu \nu}]
	\end{align*}
where 
 	\begin{equation*}
		\partial_{\mu} \eta = \frac{\partial \eta(x_1,x_2)}{\partial x_{\mu}} \Bigg\vert_{(x_1,x_2)=(\hat P_1, \hat P_2)}.
	\end{equation*}
In a similar way we can compute the commutator $[ \bar\eta (1 \otimes P+P \otimes 1) ,1 \otimes M_{\mu \nu} + M_{\mu \nu} \otimes 1]$. Using the following commutator 
 	\begin{equation*}
		[ P_{\mu} \otimes  1 + 1 \otimes  P_{\mu},1 \otimes  M_{\mu \nu}]=1 \otimes [ P_{\mu}, M_{\mu \nu}]
	\end{equation*}
and the fact that it commutes with  $ P_{\mu} \otimes  1 + 1 \otimes  P_{\mu}$, we can easily see that we have
 	\begin{align*}
		[ \bar\eta (1 \otimes P+P \otimes 1),1 \otimes  M_{\mu \nu}]&= \\
		\partial_1 \bar\eta \vert_{(1 \otimes P+P \otimes 1)}&(1 \otimes [ P_1, M_{\mu \nu}])+\partial_2 \bar\eta \vert_{(1 \otimes P+P \otimes 1)}(1 \otimes [ P_2, M_{\mu \nu}])
	\end{align*}
and an equivalent result for the commutator with $M_{\mu \nu} \otimes 1$. So we finally have, for $B=-\bar\eta (P) \otimes 1 - 1 \otimes \bar\eta(P)+\bar\eta (1 \otimes P+P \otimes 1)  + \frac{i \theta}{2}(P_1 \otimes P_2 - P_2 \otimes P_1)$ and  $C=\Delta (X)=1 \otimes M_{\mu \nu} +M_{\mu \nu} \otimes 1$, the following commutator\footnote{Do not confuse the metric $\eta_{\mu \nu}$ with the $\eta$ functions of the star product}
 	\begin{align}
		[B,C]=&-[\bar\eta (P) \otimes 1 ,M_{\mu \nu} \otimes 1]-[1 \otimes \bar\eta(P),1 \otimes M_{\mu \nu}]\nonumber \\
		&+[ \bar\eta (1 \otimes P+P \otimes 1), 1 \otimes M_{\mu \nu} +M_{\mu \nu} \otimes 1]\nonumber \\
		&+[ \frac{i \theta}{2}(P_1 \otimes P_2 - P_2 \otimes P_1), 1 \otimes M_{\mu \nu} +M_{\mu \nu} \otimes 1] \nonumber \\
		=&-(\partial_2 \bar\eta [ P_2,M_{\mu \nu}]+\partial_1 \bar\eta [P_1 ,M_{\mu \nu}]) \otimes 1\nonumber  \\
		&-1\otimes (\partial_2 \bar\eta [ P_2,M_{\mu \nu}]+\partial_1 \bar\eta [P_1 ,M_{\mu \nu}])\nonumber  \\
		&+\partial_1 \bar\eta \vert_{(1 \otimes P+P \otimes 1)}(1 \otimes [ P_1, M_{\mu \nu}])+\partial_2 \bar\eta \vert_{(1 \otimes P+P \otimes 1)}(1 \otimes [ P_2, M_{\mu \nu}]) \nonumber \\
		&+\partial_1 \bar\eta \vert_{(1 \otimes P+P \otimes 1)}( [ P_1, M_{\mu \nu}]\otimes 1)+\partial_2 \bar\eta \vert_{(1 \otimes P+P \otimes 1)}( [ P_2, M_{\mu \nu}]\otimes 1)\nonumber \\
		&-\frac{ \theta^{\alpha \beta}}{2}( (\eta_{\alpha \mu}P_{\nu}-\eta_{\alpha \nu}P_{\mu}) \otimes P_{\beta} +P_{\alpha}\otimes (\eta_{\beta \mu}P_{\nu}-\eta_{\beta \nu}P_{\mu})). \label{86}
	\end{align}
Note that, given that the commutator $ [P_{\alpha},M_{\mu \nu}]$ depends just on the momentum operators, then the commutator $[B,C]$ depends just on the momentum operators as well. From this, and the fact that the operator $B$ is also written just in terms of momentum operators, we have that $[B,[B,C]]$. This means that the sum in equation \eqref{85} stops at $n=1$, to the final expression of $\Delta_{\mathcal F} (M_{\mu \nu})$ is given by
 	\begin{equation*}
		\Delta_{\mathcal F} (M_{\mu \nu})= 1 \otimes M_{\mu \nu} +M_{\mu \nu} \otimes 1 + [B,C]
	\end{equation*}
where $ [B,C]$ is given in equation \eqref{86}. Once having this, we can compute the action of $M_{\mu \nu}$ on the monomial $x_{\rho}\star x_{\sigma}$ 
 	\begin{equation*}
		 M_{\mu \nu}(x_{\rho}\star x_{\sigma})= m_{ \star} \circ \Delta_{\mathcal F} ( M_{\mu \nu}) ( x_{\rho} \otimes x_{\sigma} )
	\end{equation*}
The action of the first two terms $1 \otimes M_{\mu \nu} +M_{\mu \nu} \otimes 1$ can be easily seen to be 
 	\begin{equation}
		(1 \otimes M_{\mu \nu} +M_{\mu \nu} \otimes 1) ( x_{\rho} \otimes x_{\sigma} )=i(\eta_{\nu \sigma}x_{\rho} \otimes x_{\mu}-\eta_{\mu \sigma} x_{\rho} \otimes x_{\nu} + \eta_{\nu \rho}x_{\mu} \otimes x_{\sigma}-\eta_{\mu \rho}x_{\nu} \otimes x_{\sigma}) \label{89}
	\end{equation}
To compute the action of $[B,C]$, let us expand the $\bar \eta$ function as $\bar \eta(x_1,x_2)=\sum_{i,j}a_{ij}x_1^i x_2^j$, where, as we said before, the terms $a_{00}$,  $a_{01}$ and  $a_{10}$ are zero. This means that all the terms in the expansion of $\partial_{\mu} \bar\eta$ have at least one partial derivative, so the action of the first two terms in the expression \eqref{86} on $x_{\rho} \otimes x_{\sigma}$ are zero. In order to compute the action of the third and forth terms on $x_{\rho} \otimes x_{\sigma}$, is is clear that we need to expand the $\bar \eta$ function up to second order so that the expansion of $\partial_{\mu} \bar\eta$ will have, at most, derivatives of order one, i.e. we need 
 	\begin{equation*}
		\bar \eta(x_1,x_2)=a_{1,1}x_1 x_2+a_{0,2} x_2^2+a_{2,0}x_1^2
	\end{equation*}
which gives
 	\begin{align*}
		\bar \partial_1\eta(x_1,x_2)&=a_{1,1} x_2+2a_{2,0}x_1\\
		\bar \partial_2\eta(x_1,x_2)&=a_{1,1}x_1+2a_{0,2} x_2
	\end{align*}
From this, we can see that the action of the third and fourth terms of the expression \eqref{86} on $x_{\rho} \otimes x_{\sigma}$ are respectively given by
 	\begin{align}
		[(a_{1,1}& P_2\otimes1+2a_{2,0}P_1\otimes1)(1 \otimes [ P_1, M_{\mu \nu}])+\nonumber \\
		&(a_{1,1} P_1\otimes1+2a_{0,2}P_2\otimes1)(1 \otimes [ P_2, M_{\mu \nu}])](x_{\rho} \otimes x_{\sigma})\nonumber\\
		=i[a_{1,1}& \eta_{2\rho}\otimes(\eta_{\nu1}\eta_{\mu \sigma}-\eta_{\mu1}\eta_{\nu \sigma})+2a_{2,0}\eta_{1\rho}\otimes(\eta_{\nu1}\eta_{\mu \sigma}-\eta_{\mu1}\eta_{\nu \sigma})+\nonumber\\
		&a_{1,1}\eta_{1\rho}\otimes(\eta_{\nu2}\eta_{\mu \sigma}-\eta_{\mu2}\eta_{\nu \sigma})+2a_{0,2}\eta_{2\rho}\otimes(\eta_{\nu2}\eta_{\mu \sigma}-\eta_{\mu2}\eta_{\nu \sigma})] \label{90}
	\end{align}
and
 	\begin{align}
		[(a_{1,1}& 1 \otimes P_2+2a_{2,0}1 \otimes P_1)( [ P_1, M_{\mu \nu}]\otimes1)+\nonumber\\
		&(a_{1,1} 1\otimes P_1+2a_{0,2}1\otimes P_2)([ P_2, M_{\mu \nu}]\otimes1)](x_{\rho} \otimes x_{\sigma})\nonumber\\
		=i[(\eta_{\nu1}&\eta_{\mu \rho}-\eta_{\mu1}\eta_{\nu \rho}) \otimes a_{1,1} \eta_{2\sigma}+ (\eta_{\nu1}\eta_{\mu \rho}-\eta_{\mu1}\eta_{\nu \rho})\otimes 2a_{2,0}\eta_{1\sigma}+\nonumber\\
		&(\eta_{\nu2}\eta_{\mu \rho}-\eta_{\mu2}\eta_{\nu \rho}) \otimes a_{1,1}\eta_{1\sigma}+(\eta_{\nu2}\eta_{\mu \rho}-\eta_{\mu2}\eta_{\nu \rho}) \otimes 2a_{0,2}\eta_{2\sigma}] \label{91}
	\end{align}
Finally, the action of the last term of \eqref{86} is given by
 	\begin{align}
		-\frac{ \theta^{\alpha \beta}}{2}&( (\eta_{\alpha \mu}P_{\nu}-\eta_{\alpha \nu}P_{\mu}) \otimes P_{\beta} +P_{\alpha}\otimes (\eta_{\beta \mu}P_{\nu}-\eta_{\beta \nu}P_{\mu}))( x_{\rho} \otimes x_{\sigma} ) \nonumber\\
		&=\frac{ \theta^{\alpha \beta}}{2}( (\eta_{\alpha \mu}\eta_{\nu \rho}-\eta_{\alpha \nu}\eta_{\mu \rho}) \otimes \eta_{\beta \sigma} +\eta_{\alpha \rho}\otimes (\eta_{\beta \mu}\eta_{\nu \sigma}-\eta_{\beta \nu}\eta_{\mu \sigma})) \nonumber \\
		&=\frac{ 1}{2}( \theta_{\mu \sigma}\eta_{\nu \rho} -\theta_{\nu \sigma}\eta_{\mu \rho} +\theta_{\rho \mu}\eta_{\nu \sigma}-\theta_{\rho \nu}\eta_{\mu \sigma}) (1\otimes1)\label{92}
	\end{align}
Combining the expressions \eqref{89}, \eqref{90}, \eqref{91} and \eqref{92} we get
 	\begin{align*}
		 M_{\mu \nu}(x_{\rho}\star x_{\sigma})= i(\eta_{\nu \sigma}x_{\rho}& \star x_{\mu}-\eta_{\mu \sigma} x_{\rho} \star x_{\nu} + \eta_{\nu \rho}x_{\mu} \star x_{\sigma}-\eta_{\mu \rho}x_{\nu} \star x_{\sigma})\\
		+\frac{ 1}{2}( \theta_{\mu \sigma}&\eta_{\nu \rho} -\theta_{\nu \sigma}\eta_{\mu \rho} +\theta_{\rho \mu}\eta_{\nu \sigma}-\theta_{\rho \nu}\eta_{\mu \sigma}) \\
		+i[a_{1,1} (&\eta_{2\rho}\eta_{\nu1}\eta_{\mu \sigma}-\eta_{2\rho}\eta_{\mu1}\eta_{\nu \sigma}+\eta_{1\rho}\eta_{\nu2}\eta_{\mu \sigma}-\eta_{1\rho}\eta_{\mu2}\eta_{\nu \sigma}\\
		+&\eta_{\nu1}\eta_{\mu \rho}\eta_{2\sigma}-\eta_{\mu1}\eta_{\nu \rho}\eta_{2\sigma}+\eta_{\nu2}\eta_{\mu \rho}\eta_{1\sigma}-\eta_{\mu2}\eta_{\nu \rho}\eta_{1\sigma})\\
		+2a_{2,0}(&\eta_{1\rho}\eta_{\nu1}\eta_{\mu \sigma}-\eta_{1\rho}\eta_{\mu1}\eta_{\nu \sigma}+\eta_{\nu1}\eta_{\mu \rho}\eta_{1\sigma}-\eta_{\mu1}\eta_{\nu \rho}\eta_{1\sigma})\nonumber\\
		+2a_{0,2}(&\eta_{2\rho}\eta_{\nu2}\eta_{\mu \sigma}-\eta_{2\rho}\eta_{\mu2}\eta_{\nu \sigma}+\eta_{\nu2}\eta_{\mu \rho}\eta_{2\sigma}-\eta_{\mu2}\eta_{\nu \rho}\eta_{2\sigma})] 
	\end{align*}
Note that if $ M_{\mu \nu}$ acts on the commutator $[x_{\rho},x_{\sigma}]_{\star}$, all the terms inside the square brackets of the previous expression will cancel with the corresponding terms of $ M_{\mu \nu}(x_{\sigma}\star x_{\rho})$, so we have
 	\begin{align*}
		 M_{\mu \nu}([x_{\rho}, x_{\sigma}]_{\star})=& i(\eta_{\nu \sigma}[x_{\rho} , x_{\mu}]_{\star}-\eta_{\mu \sigma} [x_{\rho}, x_{\nu}]_{\star} + \eta_{\nu \rho}[x_{\mu}, x_{\sigma}]_{\star}-\eta_{\mu \rho}[x_{\nu} , x_{\sigma}]_{\star})\\
		&+\frac{ 1}{2}( \theta_{\mu \sigma}\eta_{\nu \rho} -\theta_{\nu \sigma}\eta_{\mu \rho} +\theta_{\rho \mu}\eta_{\nu \sigma}-\theta_{\rho \nu}\eta_{\mu \sigma}) \\
		&-\frac{ 1}{2}( \theta_{\mu \rho}\eta_{\nu \sigma} -\theta_{\nu \rho}\eta_{\mu \sigma} +\theta_{\sigma \mu}\eta_{\nu \rho}-\theta_{\sigma \nu}\eta_{\mu \rho}) \\
		 =& i\eta_{\nu \sigma}([x_{\rho} , x_{\mu}]_{\star}-i\theta_{\rho \mu})-i\eta_{\mu \sigma} ([x_{\rho}, x_{\nu}]_{\star}-i\theta_{\rho \nu}) \\
		&+ i\eta_{\nu \rho}([x_{\mu}, x_{\sigma}]_{\star}-i\theta_{\mu \sigma})-i\eta_{\mu \rho}([x_{\nu} , x_{\sigma}]_{\star}-i\theta_{\nu \sigma})=0.
	\end{align*}


\begin{thebibliography}{9}

\addcontentsline{toc}{chapter}{Bibliography}

\bibitem{refschro}
  E. Schr\"odinger,
  ``\"Uber die Unanwendbarkeit der Geometrie im Kleinen,"
  Naturwiss. $\boldsymbol{31}$ (1934) 342-344.
\bibitem{refheis}
  W. Heisenberg,
  ``Die Grenzen der Anwendbarkeit der bisherigen Quantentheorie,"
  Z. Phys. $\boldsymbol{110}$. (1938) 251-266.
\bibitem{snyder}
  H.S Snyder,
  ``Quantized space-time,"
  Phys. Rev. $\boldsymbol{71}$. (1947) 38.
\bibitem{witten}
  N. Seiberg and E. Witten,
  ``String theory and noncommutative geometry,"
  JHEP $\boldsymbol{9909}$. (1999) 032. \href{http://arxiv.org/pdf/hep-th/9908142v3.pdf}{[arXiv:hep-th/9908142]}
\bibitem{rivasseau}
  V. Rivasseau,
  ``Why Renormalizable NonCommutative Quantum Field Theories?,"	
  \href{http://arxiv.org/pdf/0711.1748v1.pdf}{[arXiv:math-ph/0711.1748]}


\bibitem{generalstar1}
  F. Bayen, M. Flato, C. Fronsdal, A. Lichnerowicz and D. Sternheimer,
  ``Deformation Theory And Quantization. 1. Deformations Of Symplectic Structures,"
  Annals Phys. $\boldsymbol{111}$, 61 (1978).
\bibitem{generalstar2}
  F. Bayen, M. Flato, C. Fronsdal, A. Lichnerowicz and D. Sternheimer,
  ``Deformation Theory And Quantization. 2. Physical Applications,"
  Annals Phys. $\boldsymbol{111}$, 111 (1978).
\bibitem{star}
  D. Sternhaimer,
  ``Deformation quantization: Twenty years after,"
  AIP Conf.Proc. 453 (1998) 107-145.

\bibitem{gron}
  H. Gr\"onewold,
  ``On principles of quantum mechanics,"
  Physica $\boldsymbol{12}$. (1946) 405.
\bibitem{moyal}
  J. E. Moyal,
  ``Quantum mechanics as a statistical theory,"
  Proc. Cambridge Phil. Soc. $\boldsymbol{45}$, 99 (1949).
\bibitem{voros1}
  F. Bayen,
  in Group Theoretical Methods in Physics, ed. E.. Beiglbock, et. al. [Lect. Notes Phys. 94, 260 (1979)]
  \bibitem{voros2}
  A. Voros,
  ``Wentzel-Kramers-Brillouin method in the Bragmann representation,"
  Phys. Rev. A40, 6814 (1989).
\bibitem{voros3}
  M. Bordemann and S. Waldmann, 
  ``A Fedosov Star Product of Wick Type for K\"aler Manifolds," 	
  Lett. Math. Phys. 41, 243 (1997) \href{http://arxiv.org/pdf/q-alg/9605012v1.pdf}{[arXiv:q-alg/9605012]}
\bibitem{voros4}
  M. Bordemann and S. Waldmann, 
  ``Formal GNS Construction and States in Deformation Quantization," 	
  Comm. Math. Phys. 195, 549 (1998) \href{http://arxiv.org/pdf/q-alg/9607019v2.pdf}{[arXiv:q-alg/9607019]}

\bibitem{Szabo} 
R.~J.~Szabo,
  ``Quantum field theory on noncommutative spaces,''
  Phys.\ Rept.\  {\bf 378} (2003) 207
  [hep-th/0109162].

\bibitem{gallucciolizzivitalemixing}               
  S. Galluccio, F. Lizzi and P. Vitale, 
  ``Translation Invariance, Commutation Relations and Ultraviolet/Infrared Mixing," 	
  JHEP 0909:054 (2009) \href{http://arxiv.org/pdf/0907.3640v1.pdf}{[arXiv:hep-th/0907.3640]}
\bibitem{TanasaVitale}
A.~Tanasa and P.~Vitale,
  ``Curing the UV/IR mixing for field theories with translation-invariant $\star$ products,''
  Phys.\ Rev.\ D {\bf 81} (2010) 065008
  [arXiv:0912.0200 [hep-th]].
\bibitem{ArdalanSadoogi}
  F. Ardalan and N. Sadooghi, 
  ``Translational-invariant noncommutative gauge theory,"
  Phys. Rev. D 83, 025014 (2011) \href{http://arxiv.org/pdf/1008.5064v3.pdf}{[arXiv:hep-th/1008.5064]}
\bibitem{Varshovi1}
A.~A.~Varshovi,
  ``Consistent Anomalies in Translation-Invariant Noncommutative Gauge Theories,''
  J.\ Math.\ Phys.\  {\bf 53} (2012) 042303
  [arXiv:1102.4059 [hep-th]].
  \bibitem{LizziVitalereg}
  F.~Lizzi and P.~Vitale,
  ``Gauge and Poincar\'e Invariant Regularization and Hopf Symmetries,''
  Mod.\ Phys.\ Lett.\ A {\bf 27} (2012) 1250097
  [arXiv:1202.1190 [hep-th]].
\bibitem{Varshovi2}A.~A.~Varshovi,
  ``$\alpha^*$-cohomology, and classification of translation-invariant non-commutative quantum field theories,''
  J.\ Geom.\ Phys.\  {\bf 83} (2014) 53
  [arXiv:1210.0695 [math-ph]].


\bibitem{chaichian}
  M. Caichian, P. P. Kulish, K. Nishijima and A. Tureanu, 
  ``On a Lorentz-invariant Interpretation of Noncommutative Space-Time and its implications in Noncommutative QFT," 	
  Phys. Letters B. 604, 1-2 (2004) \href{http://arxiv.org/pdf/hep-th/0408069v2.pdf}{[arXiv:hep-th/0408069]}

\bibitem{lizzivitaletwist}
  S. Galluccio, F. Lizzi and P. Vitale, 
  ``Twisted Noncommutative Field Theory with the Wick-Voros and Moyal Products," 	
  Phys. Rev. D78:085007 (2008) \href{http://arxiv.org/pdf/0810.2095v1.pdf}{[arXiv:hep-th/0810.2095]}

\bibitem{lizzivitaletwistall}
  P. Aschieri, F. Lizzi and P. Vitale, 
  ``Twisting all the way: from Classical Mechanics to Quantum Fields," 	
  Phys. Rev. D77:025037 (2008) \href{http://arxiv.org/pdf/0708.3002v2.pdf}{[arXiv:hep-th/0708.3002]}
\bibitem{mixing1}
  I. Chepelev and R. Roiban, 
  ``Renormalization of quantum fileld theories on noncommutative $R^d$. 1. Scalars," 	
  JHEP 0005 (2000) 037 \href{http://arxiv.org/pdf/hep-th/9911098v4.pdf}{[arXiv:hep-th/9911098]}
\bibitem{mixing2}
  S. Minwalla, M. Van Raamsdonk and N. Seiberg,
  ``Noncommutative perturbative dynamics," 	
  JHEP 0002 (2000) 020 \href{http://arxiv.org/pdf/hep-th/9912072v2.pdf}{[arXiv:hep-th/9912072]}
\bibitem{mixing3}
  M. Hayakawa,
  ``Perturbative analysis on infrared aspects of noncommutative QED on $R^4$," 	
  Phys. Lett. B 478 (2000) 394 \href{http://arxiv.org/pdf/hep-th/9912094v3.pdf}{[arXiv:hep-th/9912094]}
\bibitem{mixing4}
  A. Matusis, L. Susskind and N. Toumbas,
  ``The IR/UV connection in the noncommutative gauge theories," 	
  JHEP 0012 (2000) 002 \href{http://arxiv.org/pdf/hep-th/0002075v2.pdf}{[arXiv:hep-th/0002075]}


\bibitem{coherent}
  K. E. Cahill and R. J. Glauber,
  ``Ordered Expansions in Boson Amplitude Operators,"
  Phys. Rev. $\boldsymbol{177}$, 5 (1969).
\bibitem{soloviev}
 M. A. Soloviev,
  ``Integral representations of the star product corresponding to the s-ordering of the creation and annihilation operators,"
  Phys. Scr. $\boldsymbol{90}$ (2015) 074008

\bibitem{IR/UVmixing}
S.~Minwalla, M.~Van Raamsdonk and N.~Seiberg,
  ``Noncommutative perturbative dynamics,''
  JHEP {\bf 0002} (2000) 020
  [hep-th/9912072].

\bibitem{peskin}
  M. Peskin and D. Schroeder,
  \emph{An Introduction to Quantum Field Theory},
  Perseus Books, Massachusetts (1995)
\bibitem{lebellac}
  M. Le Bellac,
  \emph{Quantum and Statistical Field Theory},
  Oxford University Press, New York (1991)


\bibitem{aschieri1}
  P. Aschieri, 
  ``Lectures on Hopf Algebras, Quantum Groups and Twists,"  (2007) \href{http://arxiv.org/pdf/hep-th/0703013v1.pdf}{[arXiv:hep-th/0703013]}
\bibitem{twist}
  P. Aschieri, M. Dimitrijevi\'c, F. Meyer and J. Wess,
  ``Noncommutative geometry and gravity,"
  Class. Quantum Grav.$\boldsymbol{23}$, (2006) \href{http://arxiv.org/pdf/hep-th/0510059v2.pdf}{[arXiv:hep-th/0510059]}

\bibitem{}
  F. Lizzi, M. Rivera and P. Vitale, 
  ``Green's Functions for Translation Invariant Star Products,"  (2007) \href{http://arxiv.org/pdf/1508.02575v1.pdf}{[ 	arXiv:1508.02575 [hep-th]]}

 \end{thebibliography}
\end{document}